\documentclass[final,3p,times,authoryear]{elsarticle}

\usepackage{amssymb}
\usepackage{amsmath}
\usepackage{subfigure}
\usepackage{url}
\usepackage{xcolor}

\usepackage{lineno}

\newcommand\Rey{\mbox{\textit{Re}}}  
\newcommand\Pran{\mbox{\textit{Pr}}} 

\newcommand\eg{e.g.\ }
\newcommand\ie{i.e.\ }
\newcommand{\pl}{\left(}
\newcommand{\pr}{\right)}
\newcommand{\n}{\nabla}
\newcommand{\vn}{\boldsymbol{\nabla}}
\newcommand{\vect}[1]{\boldsymbol{#1}}
\newcommand{\vel}{\mathbf{u}}
\newcommand{\vor}{\boldsymbol{\omega}}
\newcommand{\vorp}{\omega_{\phi}}

\newcommand{\pd}[2]{\frac{\partial #1}{\partial #2}}
\newcommand{\Tpert}{\Theta}
\newcommand{\Cpert}{\xi}

\newcommand{\Ek}{\mbox{\textit{Ek}}}
\newcommand{\Ra}{\mbox{\textit{Ra}}}
\newcommand{\Ro}{\mbox{\textit{Ro}}}
\newcommand{\Nu}{\mbox{\textit{Nu}}}
\newcommand{\Sc}{\mbox{\textit{Sc}}}
\newcommand{\Le}{\mbox{\textit{Le}}}

\newcommand{\asp}{\chi}
\newcommand{\Rrho}{R_{\rho}}
\newcommand{\Rrhoi}{R_{\rho}^i}
\newcommand{\smallr}{r_{\rho}}
\newcommand{\ltheo}{\mathcal{L}_\mathrm{theo}}
\newcommand{\lscale}{\mathcal{L}}
\newcommand{\Rel}{\mbox{\textit{Re}}_{\mathcal{L}}}

\newcommand{\revisiontwo}[1]{#1}
\newcommand{\revision}[1]{#1}

\journal{Physics of the Earth and Planetary Interiors}

\begin{document}

\begin{frontmatter}




\title{Influence of Rotation on Fingering Convection in \revision{a Spherical Stably Stratified Layer}}

\author[first]{Martin Gray}
\author[first]{C\'eline Guervilly\corref{cor1}}
\author[first]{Graeme R. Sarson}
\affiliation[first]{organization={School of Mathematics, Statistics and Physics, Newcastle University},city={Newcastle Upon Tyne},postcode={NE17RU},country={UK}}
\cortext[cor1]{corresponding author: \url{celine.guervilly@ncl.ac.uk}}

\begin{abstract}
Stably stratified layers are thought to develop at the top of the liquid metallic cores of many terrestrial planets. We consider the case where the thermal gradient is stable but the compositional gradient is unstable, a situation particularly relevant to Mercury. The strong contrast between molecular diffusivities of temperature and composition leads to fingering convection. We investigate this process using hydrodynamical simulations in a rotating spherical shell, systematically varying the stratification strength $N$ relative to the rotation rate $\Omega$. In all regimes, the primary fingering mode forms narrow, elongated structures that shift orientation from the rotation axis to the direction of gravity as $N^2/\Omega^2$ exceeds 10. The fingers remain laminar, with transverse scales proportional to thermal stratification but independent of rotation. Fingering convection also drives secondary large-scale flows across most of the explored parameter space, producing diverse dynamics including zonal flows, hemispherical convection, axisymmetric poloidal bands, finger clusters, and toroidal gyres.
In the rapidly-rotating regime, laterally inhomogeneous mixing generates zonal flows in thermo-compositional wind balance; zonal flow direction and amplitude depend on $N^2/\Omega^2$, with amplitude weakening for strong stratification $N^2/\Omega^2>10$. 
In the intermediate regime ($N^2/\Omega^2 \sim 1$), axisymmetric or spiraling poloidal bands emerge within the tangent cylinder, gradually overtaking the primary fingers. 
For stronger stratification, finger clusters and weak, large-scale density anomalies surrounded by toroidal gyres form in the upper domain. These diverse large-scale flows may interact with the dynamo-generated magnetic field in the deeper core, potentially influencing surface magnetic fields.
\end{abstract}




\end{frontmatter}




\section{Introduction}
\label{sec:intro}

\revision{
Many terrestrial planets are expected to host a stably stratified layer at the top of their liquid metallic cores as a consequence of thermal and compositional evolution. Thermal stratification can arise when the heat flux across the core-mantle boundary becomes locally sub-adiabatic \citep[\eg][]{Labrosse2015,Greenwood2021} while compositional stratification may result from the accumulation of light elements near the top of the core \citep[\eg][]{Buffett2010,Bouffard2019}. 
For the Earth, seismological and geomagnetic observations have been interpreted as evidence for such a stable layer in the uppermost outer core \citep{Helffrich2010,Buffett2014}, although its existence, properties and extent remain debated \citep{Lesur2015,Huguet2018,Irving2018}. Models of other planetary bodies, including Mercury and Ganymede, similarly suggest thermally stratified regions within their cores \citep{Hauck2004,Dumberry2015,Davies2024,Knibbe2025}. These stratified layers are often assumed to inhibit radial convective motions while permitting large-scale differential rotation, which may have important implications for the morphology of planetary magnetic fields. For example, Mercury’s magnetic field is weak, dominantly dipolar, and unusually axisymmetric \citep{Anderson2012}, features that have been linked to the filtering effect of a stable layer on the dynamo-generated field \citep{Christensen2006,Christensen2008}.}

\revision{
However, the common assumption that stable density gradients largely suppress vertical motions ignores an important aspect of stably stratified fluid dynamics. If one source of buoyancy (thermal or compositional) is destabilising within the stratified layer, the system can become susceptible to double-diffusive instabilities \citep{Radko2013}. In particular, a configuration in which the temperature gradient is stabilising while the compositional gradient is destabilising (as is plausible for the upper cores of Mercury and Ganymede) gives rise to fingering convection, a type of double-diffusive convection in which the primary motions take the form of vertical plumes. 
Consequently, the assumption that radial convective flows are largely suppressed in the stable layer may not hold, raising questions about the influence of stratification on the observed planetary magnetic fields. The presence of fingering convection could have significant effects on the morphology of the magnetic fields. Yet, the interaction between fingering convection and magnetic fields remains largely unexplored \citep{Harrington2019,Fraser2024}.
}

Fingering convection occurs when a stabilising thermal gradient is opposed by an unstable compositional gradient, and where thermal diffusivity greatly exceeds compositional diffusivity \citep{Stern1960}.
The potential energy linked to the destabilising composition is released through the rapid molecular diffusion of temperature. This process drives vertical motions that can greatly enhance the transport of heat and chemical composition in otherwise stable layers \citep{Traxler2011b,Stellmach2011}. Fingering convection has been widely explored in fields such as oceanography \citep{Schmitt1994,Kunze2003}, magma dynamics \citep{Sparks1984,Hansen1990}, and stellar interior dynamics \citep{Garaud2021}. The potential for fingering convection to operate in the metallic cores of planets has recently received increasing attention \citep {Manglik2010,Net2012,Bouffard2017,Monville2019,Silva2019,Mather2021,Guervilly2022,Tassin2024}. 
A key distinction between these different natural systems lies in the value of the Prandtl number  ($\Pran$), which quantifies the ratio of kinematic viscosity to thermal diffusivity. While oceans and magmas typically have $\Pran > 1$, liquid metals such as those found in planetary cores exhibit lower values, on the order of $\Pran \sim 0.1$. Studies of fingering convection in stellar interiors, where $\Pran \ll 1$, have demonstrated that the Prandtl number plays a critical role in controlling the dynamics \citep{Garaud2018}, thereby limiting the direct applicability of results across different natural systems. One notable example is the near-absence of thermo-compositional staircases at low Prandtl numbers \citep{Traxler2011, Brown2013}. These staircases, which consist of well-mixed layers separated by sharp stratified interfaces, are known to substantially enhance mixing efficiency when present at high Prandtl numbers \citep[\eg][]{Stellmach2011}.

\revisiontwo{The motivation of this work is to characterise the dynamics of fingering convection under conditions relevant to metallic cores. As a first step in this direction, we use a simplified spherical shell model, where the dynamics of the stable layer is studied in isolation from the deeper convective layer. The study focuses on thick stable layers, which are expected to be particularly suitable for Mercury. Thermally stratified regions in liquid outer cores are also relevant for the Earth \citep[\eg][]{Greenwood2021} and Ganymede \citep{Hauck2006}. }
This work builds on recent studies by \citet{Guervilly2022} and \citet{Tassin2024}, both of which explore fingering convection in spherical shell geometries across wide regions of parameter space. \citet{Tassin2024} investigates the non-rotating regime, while \citet{Guervilly2022} considers the regime where rotational effects dominate over stratification (a regime that we will quantify below). In the latter case, at low forcing (\ie small Rayleigh numbers), the flow organises into tall, columnar structures aligned with the rotation axis, with azimuthal size comparable to the layer thickness \citep{Monville2019,Mather2021}. As the forcing increases, these structures become increasingly sheet-like and thin in azimuth while remaining axially elongated. Radial motions remain largely laminar, and strong equatorially-symmetric zonal (\ie axisymmetric and azimuthal) flows develop, eventually dominating the non-axisymmetric flow. In contrast, in the absence of rotation, \citet{Tassin2024} finds that the fingers align radially. \citeauthor{Tassin2024} further examines how thermal and compositional fluxes, velocities, and finger sizes scale with key dimensionless parameters, deriving new scaling laws. Additionally, they identify a secondary instability characterised by the formation of large-scale toroidal jets that distort the fingering structures and induce relaxation oscillations. Given the uncertainty surrounding the stratification strength in Mercury’s stable layer, the present study aims to bridge the gap between the strongly-stratified, non-rotating regime of \citet{Tassin2024} and the weakly-stratified, rapidly-rotating regime of \citet{Guervilly2022}. We focus specifically on identifying the conditions under which the transition between these regimes occurs. 
As a first step in addressing this complex problem, magnetic fields are neglected in the present model.

The paper is structured as follows. Section~\ref{sec:model} presents the mathematical and computational formulation of the model. Section~\ref{sec:results} details the results, including the dynamics across weak, intermediate, and rapid rotation regimes, the emergence of differential rotation (\ie zonal flows), and the efficiency of convective transport. Concluding remarks are provided in Section~\ref{sec:conclusion}.

\section{Model}
\label{sec:model}

\subsection{Governing equations}
\label{sec:equations}

We model a stably stratified layer of Boussinesq fluid confined within a spherical shell. The inner boundary, at radius $R_i$, represents the interface with the deeper convective core, while the outer boundary, at radius $R_o$, corresponds to the core-mantle boundary. The thickness of the stable layer at the top of Mercury's core is not well constrained by thermal evolution models. Here we model a thick layer of depth $800$\,km, which is at the upper end of the proposed range \citep{Wardinski2021,Davies2024}. This corresponds to an aspect ratio of $\asp = R_i/R_o = 0.6$. 

The fluid layer rotates about the $z$-axis with rotation rate $\Omega$. Gravity is directed radially inward and varies linearly with spherical radius $r$, given by $\vect{g} = -g_0 r/R_o \vect{e}_r$ where $g_0$ is a constant. The fluid has constant viscosity $\nu$, thermal diffusivity $\kappa_t$ and compositional diffusivity with $\kappa_t>\nu>\kappa_c$.  

The density $\rho$ depends linearly on the temperature $T$ and the composition $C$ (\ie the concentration in light elements) of the Boussinesq fluid:
\begin{equation}
	\frac{\rho}{\rho_m} =  1- \alpha_t (T-T_m) - \alpha_c(C-C_m),
\label{eq:state}
\end{equation}
where $\alpha_t$ is the coefficient of thermal expansion, and $\alpha_c$ the compositional analogue to $\alpha_t$, with constant $\alpha_t, \alpha_c>0$.
The subscript $m$ denotes a constant mean value. 
The temperature and composition are decomposed into a constant mean, a static profile (denoted by the subscript $s$) and a perturbation:
\begin{eqnarray}
	&& T(r, \theta, \phi, t) = T_m+T_s(r) + \Tpert(r, \theta, \phi, t), \quad 
 \\
   && C(r, \theta, \phi, t) = C_m+C_s(r) +\Cpert(r, \theta, \phi, t),
\end{eqnarray}
in spherical polar coordinates $(r, \theta, \phi)$ and where $t$ is time.

The equations are nondimensionalised using the layer thickness $D=R_o-R_i$ as a unit for lengths, $D^2/\nu$ for times, $\nu^2/\alpha_t g_o D^3$ for temperature, and  $\nu^2/\alpha_c g_o D^3$ for composition.
The dimensionless inner and outer radii are thus $r_i=R_i/D=1.5$ and $r_o=R_o/D=2.5$.
The dimensionless governing equations are
\begin{eqnarray}
	\pd{\vel}{t} + (\vel \cdot \vn )\vel + \frac{2}{\Ek}\vect{e}_z \times \vel 
    &=& -\vn p + \vn^2 \vel + \pl \Tpert + \Cpert \pr  \frac{r}{r_o} \vect{e}_r,
	\label{eq:NS}
    \\
 	\pd{\Tpert}{t} + \vel \cdot \vn \Tpert  + u_r \frac{\textrm{d} T_s}{\textrm{d}r}  &=&  \frac{1}{\Pran} \n^2 \Tpert, \label{eq:Tpert}
	\\
	\pd{\Cpert}{t} + \vel \cdot \vn \Cpert + u_r \frac{\textrm{d} C_s}{\textrm{d}r}  &=&  \frac{1}{\Sc} \n^2 \Cpert, \label{eq:Cpert}
\end{eqnarray}
where $\vel=(u_r,u_{\theta},u_{\phi})$ is the solenoidal velocity and $p$ the pressure. 
The dimensionless numbers are the Ekman, Prandtl, and Schmidt numbers defined respectively as
\begin{equation}
	\Ek = \frac{\nu}{\Omega D^2},~~~ \Pran=\frac{\nu}{\kappa_t},~~~ 
    \Sc=\frac{\nu}{\kappa_c}.
\end{equation}

The radial gradients of the static fields are obtained by solving the diffusion equations in the absence of motions. 
At the inner boundary, the static temperature and composition gradients are imposed:
\begin{eqnarray}
	T_s' (r_i) = \left.  \frac{\textrm{d} T_s}{\textrm{d}r} \right|_{r_i} =  - \frac{\Ra_t}{\Pran} \frac{1}{r_i}, ~~~
	C_s' (r_i) = \left.  \frac{\textrm{d} C_s}{\textrm{d}r} \right|_{r_i} = - \frac{Ra_c}{\Sc}  \frac{1}{r_i} ,
\end{eqnarray}
where the Rayleigh numbers are dimensionless numbers defined as
\begin{equation}
	\Ra_t = \frac{\alpha_t g_o q_i R_i D^3}{\nu \kappa_t}, \quad \Ra_c = \frac{\alpha_c g_o f_i R_i D^3}{\nu \kappa_c},
\end{equation}
where $q_i\propto -T_s'(r_i)$ and $f_i\propto -C_s'(r_i)$ are the dimensional static temperature and composition fluxes at the inner boundary. 
In fingering convection, the static temperature gradient through the layer is stable while the static compositional gradient is unstable, so $q_i<0$ and $f_i>0$, and so $\Ra_t<0$ and $\Ra_c>0$.
At the core-mantle boundary, we assume that there is no flux of light elements to/from the mantle, hence $C_s'(r_o)=0$. We therefore include a sink term in the diffusion equation for $C_s$ to compensate the inflow of light elements at the inner boundary and achieve a steady state (see further detail in \citet{Guervilly2022}). For the static temperature, the outward heat flux integrated over the spherical surface at $r_o$ equals the incoming heat flux integrated over the spherical surface at $r_i$ to ensure a steady state, so no source term are included in the diffusion equation for $T_s$.
The gradient of the static fields in the domain are thus
\begin{eqnarray}
	&&T_s'= \frac{\textrm{d} T_s}{\textrm{d}r} =  - \frac{\Ra_t}{\Pran} \frac{r_i}{r^2},
  \label{eq:dTs}
	\\
	&&C_s'=\frac{\textrm{d} C_s}{\textrm{d} r} = \frac{Ra_c}{\Sc(1-\asp^3)}  \pl \frac{r_i }{r_o^3} r - \frac{r_i}{r^2} \pr.
 \label{eq:dCs}
\end{eqnarray}

The boundary conditions for the perturbations are zero radial fluxes at $r=r_i$ and $r=r_o$,
\begin{linenomath*}
\begin{eqnarray}
	\left. \pd{\Tpert}{r} \right|_{r_i} = \left. \pd{\Cpert}{r} \right|_{r_i} = \left. \pd{\Tpert}{r} \right|_{r_o} = \left. \pd{\Cpert}{r} \right|_{r_o} = 0.
\label{eq:BC_TC}
\end{eqnarray}
\end{linenomath*}

We neglect any dynamical interaction with the deep convective core at $r=r_i$ for simplicity. We therefore use impenetrable boundary conditions at $r=r_i$ and $r=r_o$. We use stress-free boundary conditions at $r=r_i$ to model the interface with the convective liquid core and no-slip boundary conditions at $r=r_o$ to model the rigid core-mantle boundary.
\revision{The choice of impenetrable and stress-free boundary condition at $r=r_i$ does not capture the interaction between a stably stratified layer and an underlying convective region. This choice is made primarily for numerical convenience, enabling a systematic exploration of parameter space. This simplification also allows us to study the intrinsic dynamics of the stably stratified layer in isolation.}

The diffusivity coefficients are held constant throughout this study. Following \citet{Monville2019, Guervilly2022}, we adopt a Prandtl number of $\Pran = 0.3$ and a Schmidt number of $\Sc = 3$, which yields a Lewis number $\Le = \kappa_t / \kappa_c = \Sc / \Pran = 10$ for all simulations. Although this value is lower than that expected in planetary cores ($\Le = \mathcal{O}(10^3)$), it remains computationally tractable and allows for a meaningful exploration of double-diffusive fingering instabilities within a relevant parameter regime as described below.

\subsection{Stratification and domain of fingering convection}

The stratification of the layer measures the stability of the fluid to vertical displacements. It is quantified by the buoyancy frequency $N$, also called the Brunt-V\"ais\"al\"a frequency, which corresponds to the frequency of oscillations of a vertically-displaced fluid parcel under the restoring force of gravity. 
$N^2$ is proportional to minus the radial density gradient, so in a stably stratified layer, where the density decreases with radius,  $N^2>0$. The system is unstably stratified when $N^2<0$. We measure the buoyancy frequency from the static profiles~\eqref{eq:dTs}-\eqref{eq:dCs}, which gives in dimensionless units, 
\begin{equation}
	\tilde{N}^2 = \frac{N^2}{\nu^2/D^4} = \frac{r}{r_o} \left( T_s' + C_s'\right). 
\end{equation} 
For our choice of static gradients, $N^2$ varies slightly with radius, with the stratification being strongest at the bottom of the layer and $N^2(r_o)=0.85 N^2(r_i)$.  In the rest of the paper, values of $N^2$ will always be quoted at $r=r_i$. 

In a rotating domain, it is convenient to measure $N$ relative to the rotation rate $\Omega$:
\begin{equation}
    \frac{N^2}{\Omega^2} = - \frac{\Ek^2}{r_0}\left(\frac{\Ra_t}{\Pran} + \frac{\Ra_c}{\Sc} \right) 
    = \frac{\Ek^2}{r_0} \frac{\Ra_c}{\Sc} \left(\Rrhoi -1\right),
    \label{eq:N2}
\end{equation}
where the density ratio $ \Rrho$ is 
\begin{equation}
    \Rrho=\frac{|T_s'|}{|C_s'|}, ~~ \Rrhoi=\Rrho(r_i)= \frac{|\Ra_t| \Le}{\Ra_c}.
\end{equation} 

Double-diffusive instabilities are commonly characterised by the density ratio, which delineates the conditions under which they may arise.
In non-rotating, planar layers with uniform static gradients, the system becomes linearly unstable to the double-diffusive fingering instabilities when 
\begin{equation}
	1<R_{\rho}<\Le,
	\label{eq:range}
\end{equation}
assuming $\Le\gg1$ and sufficiently high Rayleigh numbers \citep{Stern1960,Baines1969}.
Values of $R_{\rho}<1$ correspond to unstably stratified systems with $N^2<0$.
Rotational effects do not significantly alter this critical range for large $\Ra_c$ \citep{Sengupta2018,Monville2019}.
However, in spherical geometry with rotation, large-scale double-diffusive modes with low azimuthal wavenumber can be preferred to small-scale fingers near the stability boundaries, especially for small $\Ra_c$  \citep{Silva2019,Monville2019,Guervilly2022}. 
Fingering convection can also be preferred for $R_{\rho}<1$ \citep{Hage2010,Schmitt2011,Kellner2014,Yang2015}.
Since $\Le \gg 1$ in planetary cores, the parameter regime prone to fingering instability is therefore broad, suggesting that fingering convection may play a role across a wide range of thermal and compositional stratifications.

In our study, $R_{\rho}$ increases with radius (see Figure 2 in \citet{Guervilly2022}). At the bottom of the domain, the unstable range~\eqref{eq:range} corresponds to
\begin{equation}
	\Ra_c/\Le < |\Ra_t| < \Ra_c.
	\label{eq:range2}
\end{equation}
The density ratio goes to infinity at the outer boundary as $C_s'$ is zero there. $\Rrho$ crosses the theoretical onset of fingering convection for $\Rrho(r=r_s)=\Le$. This implies that, while the bottom of the domain might be prone to fingering instabilities, the top of the domain might not be. 
As the radial profile of $R_{\rho}$ depends on $|\Ra_t|/\Ra_c$ (and not on $\Ra_t$ or $\Ra_c$ individually),  fixing this ratio leaves radius $r_s$ unchanged. 
In this study, we fix \mbox{$\Ra_t=-\Ra_c/3$} \ie $\Rrhoi=\Le/3$. 
This implies that $r_s=2.26$, so we would expect fingering convection to occur in the inner $76\%$ of the radial domain at large Rayleigh numbers.

\subsection{Numerical method}
All the simulations are performed with the open-source code XSHELLS \citep{Schaeffer2013,Schaeffer2017}.
XSHELLS is a C++ pseudo-spectral code that solves the governing equations \eqref{eq:NS}-\eqref{eq:Cpert} in a 3D spherical geometry. 
To ensure that the velocity field is solenoidal, $\vel$ is decomposed into poloidal and toroidal scalars, $U_p$ and $U_t$,
\[ \vel = \nabla \times \nabla \revision{\times} ( U_p \vect{r}) + \nabla \times (U_t \vect{r}), \]
where $\vect{r}$ is the position vector. 
All the scalars ($U_p$, $U_t$, $\Tpert$ and $\Cpert$)
are expanded in spherical harmonics $Y_l^m(\theta,\phi)$ of degree $l$ and order $m$, which are truncated at $L_{max}$ and $M_{max}$ respectively.
In the radial direction, the code uses a second-order finite difference scheme with $N_r$ points. 
The code uses a second-order time-stepping scheme with an implicit treatment of the diffusive terms and explicit treatment of the non-linear terms. 
Details of the numerical resolution used for each simulation are provided in \ref{sec:appA}.

\section{Results}
\label{sec:results}

\subsection{Overview}
\label{sec:overview}

\subsubsection{Parameter survey}

\begin{figure}
	\centering 
	\includegraphics[width=0.6\textwidth]{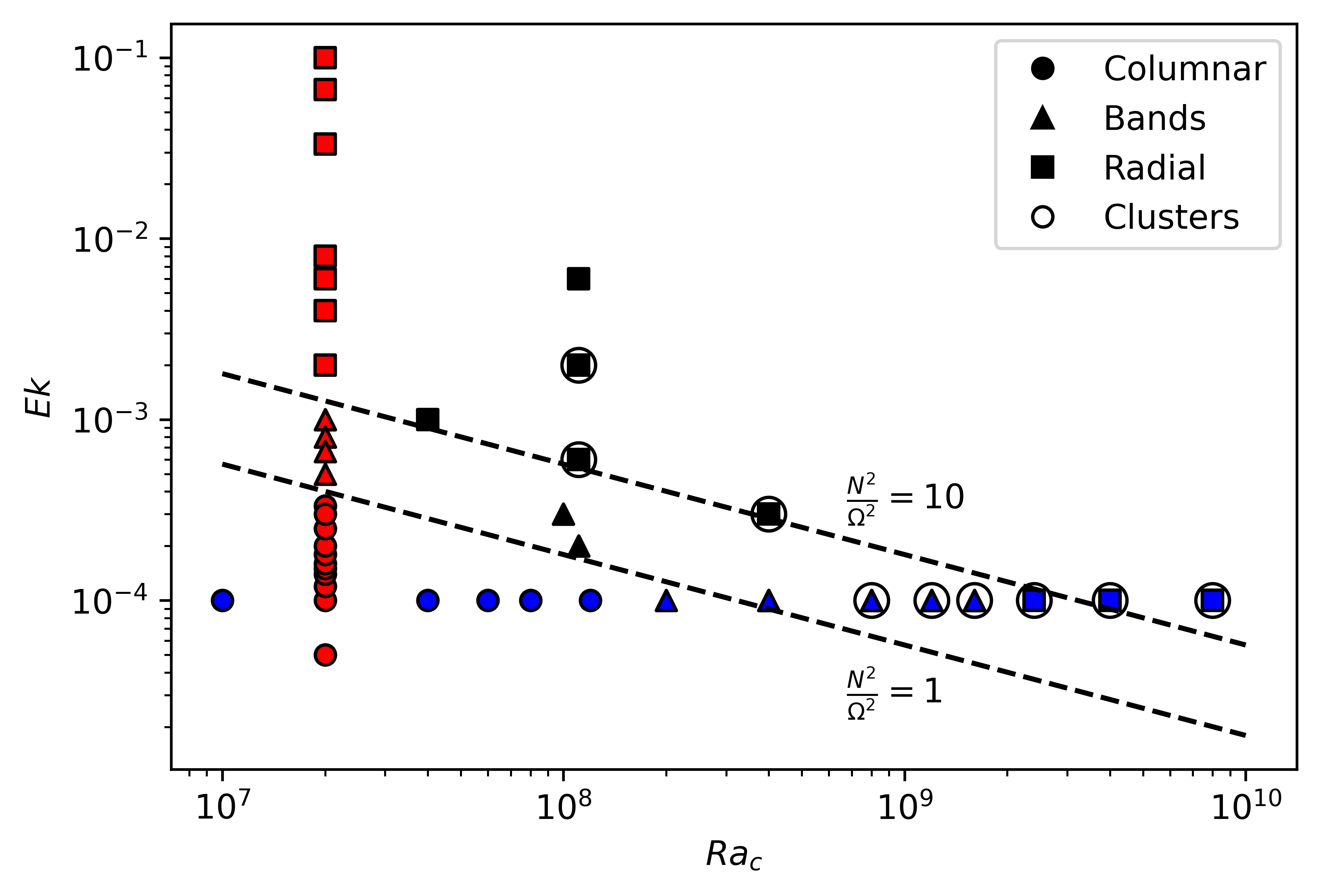}	
	\caption{Simulations in the parameter space $\Ek-\Ra_c$. All the simulations were performed with $\Le=10$, $\Pran=0.3$ and $\Rrhoi=\Le/3$. The lower (upper) dashed line represents \mbox{$N^2/\Omega^2=1$} (\mbox{$N^2/\Omega^2=10$} respectively). Series 1 simulations are represented in red and Series 2 simulations in blue. Columnar/radial/bands refer to the structure of the radial flows
 in the rapidly-rotating/weakly-rotating/intermediate regimes. The location of the finger clusters (\S\ref{sec:cluster}) is indicated by large open circles.} 
	\label{fig:Ek_vs_Ra}
\end{figure}

We first give an overview of the results before describing the flow
in the different dynamical regimes in later sections. 
All the simulations were run at a constant density ratio $\Rrhoi=\Le/3$.
As shown in Figure~\ref{fig:Ek_vs_Ra}, the ratio $N^2/\Omega^2$ is varied in two ways: 
(1) varying Ekman number at fixed Rayleigh number ($\Ra_c = 2\times10^7$) (Series 1); 
(2) varying Rayleigh numbers (\ie varying $\Ra_c$ and $\Ra_t$ by the same factor to maintain constant $\Rrhoi$)
at fixed Ekman number ($\Ek = 10^{-4}$) (Series 2). 
Tables~\ref{tab:Series1} and \ref{tab:Series2} in Appendix A list the input parameters and numerical resolutions of all the simulations. 
The range of the dynamical lengthscales increases with the Rayleigh numbers, implying that the exploration of the 
regime $N^2/\Omega^2\gg1$ is more computationally demanding in Series 2 and, consequently, restricted to 
$N^2/\Omega^2\lesssim 25$.
\revision{Simulations with increasing $N^2/\Omega^2$ in Series 1 correspond to $\Ek\to1$, so the weakening of rotation relative to stratification is accompanied by enhanced viscous effects (\eg cases with $N^2/\Omega^2>10$ correspond to $\Ek>10^{-3}$). While these simulations offer useful insights to complement the results from Series 2, their interpretation should be approached carefully in the context of planetary core applications where $\Ek\ll1$.}

\begin{figure}
	\centering 
	\includegraphics[width=0.6\textwidth]{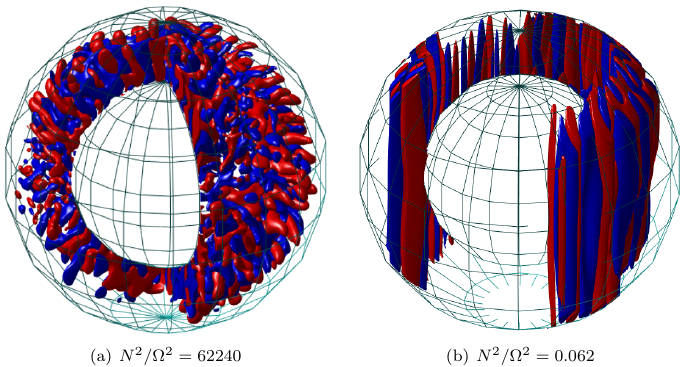}	
	\caption{Isosurfaces of the radial velocity for two representative cases of (a) $N^2/\Omega^2\gg1$ ($Ek=10^{-1}$ in Series 1) and (b) $N^2/\Omega^2\ll 1$ ($Ek=10^{-4}$ in Series 1). Red (blue) indicates positive (negative) values of $u_r$ and the isosurfaces corresponds to 10\% of the maximum values of $u_r$ in a given snapshot.}
	\label{fig:Isosurface}
\end{figure}

For convenience, we define the weakly-rotating regime as $N^2/\Omega^2>\mathcal{O}(1)$, understood to be relative to the strength of the stratification. 
Conversely, we refer to $N^2/\Omega^2<\mathcal{O}(1)$ as the rapidly-rotating regime.
\revision{In single-component rotating convection, the convective Rossby number is often used as a control parameter that determine regime transitions \citep[\eg][]{Aurnou2020}. Here we base its definition on the control parameters of compositional buoyancy, $\Ro_c=\sqrt{\Ra_c\Ek^2/\Sc}$. From equation~\eqref{eq:N2}, $N^2/\Omega^2$ is directly related to the convective Rossby number for fixed density ratio. Here $\Rrhoi=3.33$, so $N^2/\Omega^2\approx \Ro_c^2$; Values of $\Ro_c$ are given in Tables~\ref{tab:Series1} and \ref{tab:Series2}.}

Two cases representative of $N^2/\Omega^2\gg1$ and $N^2/\Omega^2\ll 1$ are illustrated in Figure~\ref{fig:Isosurface}, which show snapshots of the isosurfaces of the radial velocity.
Comparison of the two cases show one immediate difference: whilst the structure of the radial velocity in 
the weakly-rotating case is essentially aligned with gravity (``radial'' fingers), 
it is aligned with the rotation axis in the rapidly-rotating case (``columnar'' fingers).
In both cases, the fingers do not occupy the whole volume of the sphere: the density ratio increases with radius in our system
so the fingers develop preferentially close to the inner boundary. 
Additionally, fingering convection within the tangent cylinder is inhibited for rapid rotation, and only sets in at higher values of $N^2/\Omega^2$. 
This is due to the detrimental effect of rotation on the buoyancy-driven instability, which is particularly pronounced when the rotation axis is aligned with gravity, similar to the behaviour observed in single-component overturning convection \citep[\eg][]{Gastine2023}. 
Fingering convection can occur inside the tangent cylinder for $N^2/\Omega^2<1$ as shown
in Section~\S\ref{sec:rapid}.

\subsubsection{Finger scale}

\begin{figure}
	\centering 
	\includegraphics[width=\textwidth]{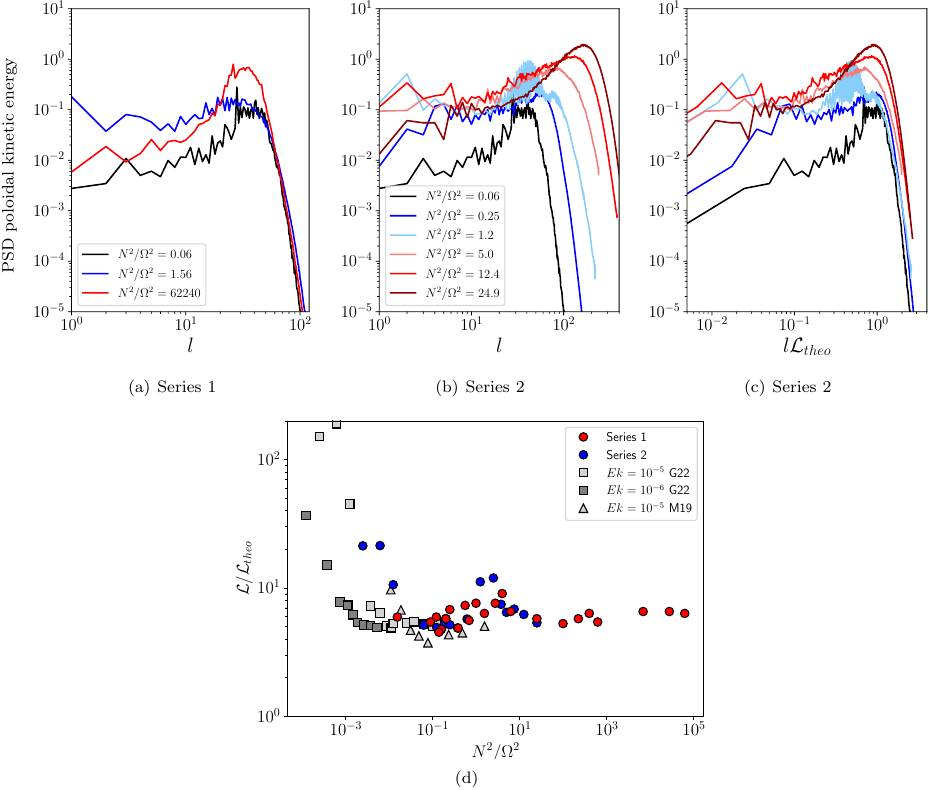}
	\caption{\revision{Top row: Power Spectral Density (PSD)} of the poloidal kinetic energy as a function of the spherical harmonics degree $l$ 
    for selected cases in Series 1 and 2 in (a)-(b).
    The spectra are calculated from a snapshot and averaged in radius excluding the boundary layers.
    In (c), the degree $l$ is normalised by $\ltheo= (|\Ra_t|/r_o)^{-1/4}$.
    $N^2/\Omega^2=0.06$ has the same parameters in both Series.
    \revision{Bottom row: Variation of the lengthscale measured from the spectral peak $\lscale$ normalised by $\ltheo$ as a function of $N^2/\Omega^2$. Data from \citet{Monville2019} (M19) and \citet{Guervilly2022} (G22) are included for smaller $\Ek$.}}
	\label{fig:spectra}
\end{figure}

In the weakly- and rapidly-rotating regimes, the fingers have a specific lengthscale in the transverse
direction (\ie the direction tangential to the rotation axis for rapid rotation and to the 
radial direction for weak rotation). 
In non-rotating plane layer, the fingering instability develops as elongated fingers with a transverse 
lengthscale that follows the dimensionless ``Stern scale'' $\ltheo = (|\Ra_t|/r_o)^{-1/4}$ for large Rayleigh numbers
and when $1\ll\Rrho\ll\Le$ \citep{Stern1960,Schmitt1979}.
In spherical shell simulations without rotation, \citet{Tassin2024} found that the observed transverse finger scale closely follows the Stern scale's dependence on $\Ra_t$, although it also exhibits sensitivity to the density ratio. In the rapidly-rotating regime, \citet{Guervilly2022} demonstrated that, at a fixed density ratio and sufficiently large Rayleigh numbers, the transverse finger scale remains well described by the Stern scale across varying Ekman numbers. This suggests that, for $N^2/\Omega^2<1$, rotation does not significantly influence the transverse finger scale.

To measure a typical transverse finger scale, we estimate the most energetic scale of the poloidal flow from the power spectra of
the poloidal kinetic energy in Series 1 and 2 cases (Figure~\ref{fig:spectra}). 
\revision{The power spectra density of the kinetic energy is used to identify the dominant energy-carrying horizontal lengthscales of the flow and the location of the spectral peak might reflect transitions between different dynamical regimes.}
In Series 1, where $\Ra_t$ is held constant, the spectra exhibit a well-defined peak for both $N^2/\Omega^2<1$ and 
$N^2/\Omega^2>1$. In both cases, the peak occurs at approximately the same wavenumber, consistent with the Stern scale, indicating that it is largely insensitive to variations in $\Ek$. 
The most noticeable difference is that length scales at or above the Stern scale are more energetic for $N^2/\Omega^2>1$.
In Series 2, where $\Ra_t$ increases, a well-defined spectral peak is again observed for both 
$N^2/\Omega^2<1$ and $N^2/\Omega^2>1$. As $N^2/\Omega^2$ increases, the flow becomes more energetic across all scales.
When the spherical harmonics degree $l$ is normalised by $\ltheo$ on the $x$-axis (Figure~\ref{fig:spectra}c), the spectral peaks align, indicating consistency with the predicted scaling.
The intermediate regime \revision{($N^2/\Omega^2\approx 1$)} is visibly different:  in Series 1, the spectrum displays a flat slope at low wavenumbers up to the Stern scale; 
in Series 2, two separate spectral peaks emerge: one at the Stern scale and another at approximately half that wavenumber. 

\revision{To quantify the departure from the Stern scale, we measure the characteristic lengthscale of the most energetic poloidal flow by identifying the spectral peak (spherical harmonic degree $l_{\mathrm{peak}}$) in each simulation. This peak is computed from the $l$ spectra averaged radially, excluding the boundary layers. The corresponding length scale, $\lscale = \pi r_i / l_{\mathrm{peak}}$, is shown in Figure~\ref{fig:spectra}(d), where it is normalised by $\ltheo$. We also include data from \citet{Monville2019} and \citet{Guervilly2022} at lower Ekman numbers (and same $\Pran$ and $\Le$) for comparison. Across all studies, at small Rayleigh numbers ($\Ra_c \Ek \lesssim 10^3$–$10^4$), the columnar flows exhibit a large azimuthal extent, comparable to the layer thickness \citep[as also observed by][]{Mather2021}. As the Rayleigh number increases, these structures evolve into narrow, sheet-like fingers (see Figure~\ref{fig:Isosurface}(b)), whose typical transverse scale progressively approaches the Stern scale and is well approximated by $\lscale = 5\ltheo$, independent of $\Ek$.
This behaviour confirms that, at sufficiently large Rayleigh numbers and for a fixed density ratio, the transverse scale of both columnar and radial fingers is governed by the Stern scale. In contrast, the intermediate regime is characterised by the emergence of larger scales, which is particularly evident in the data points from Series 2.}
Section~\S\ref{sec:intermediate} will explore this novel feature.

\subsubsection{Scaling of the radial velocity}

\begin{figure}
	\centering 
    	\includegraphics[width=0.5\textwidth]{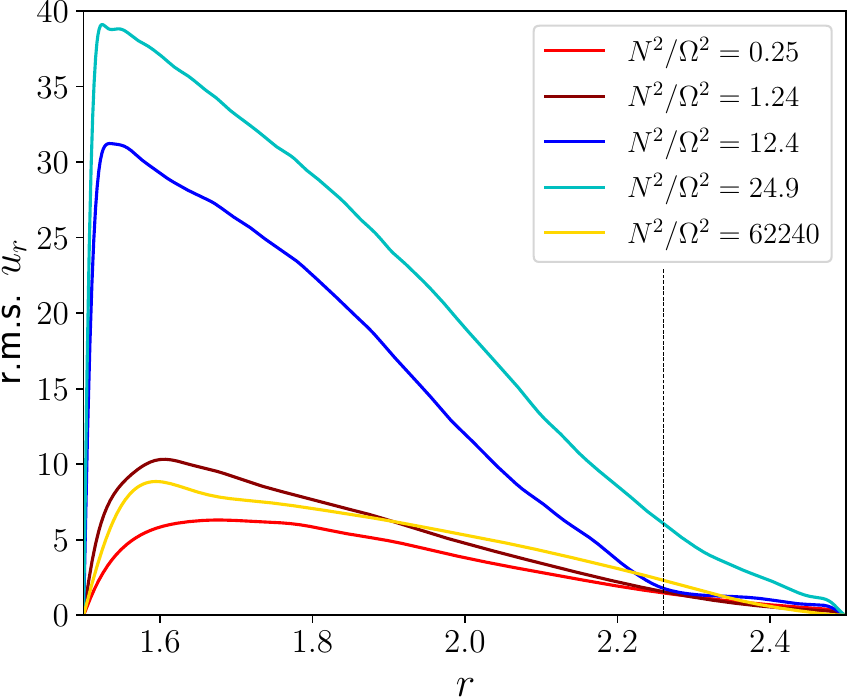}
	\caption{Radial profile of the r.m.s. radial velocity for selected simulations in Series 2 ($0.25 \leq N^2/\Omega^2\leq 24.9$) and Series 1 ($N^2/\Omega^2=62240$). The vertical dashed line corresponds to the radius $r_s$ where $\Rrho(r=r_s)=\Le$.}
	\label{fig:profile_ur}
\end{figure}

To evaluate how the radial variation of $\Rrho$ influences the vigour of fingering convection throughout the layer, Figure~\ref{fig:profile_ur} shows the radial profile of the root mean square (r.m.s.) radial velocity for increasing values of $N^2/\Omega^2$. 
\revision{The radial profiles are used to assess the radial localisation of the kinetic energy, which indicates how the active regions of the dynamics shift with control parameters. }
In all cases, radial motions are strongest near the bottom of the layer, just outside the lower boundary layer, and decrease monotonically with radius. Only weak radial flows are observed to cross the upper region beyond $r>r_s$ (where $\Rrho(r=r_s)=\Le$).

\begin{figure}
	\centering 
    	\includegraphics[width=0.5\textwidth]{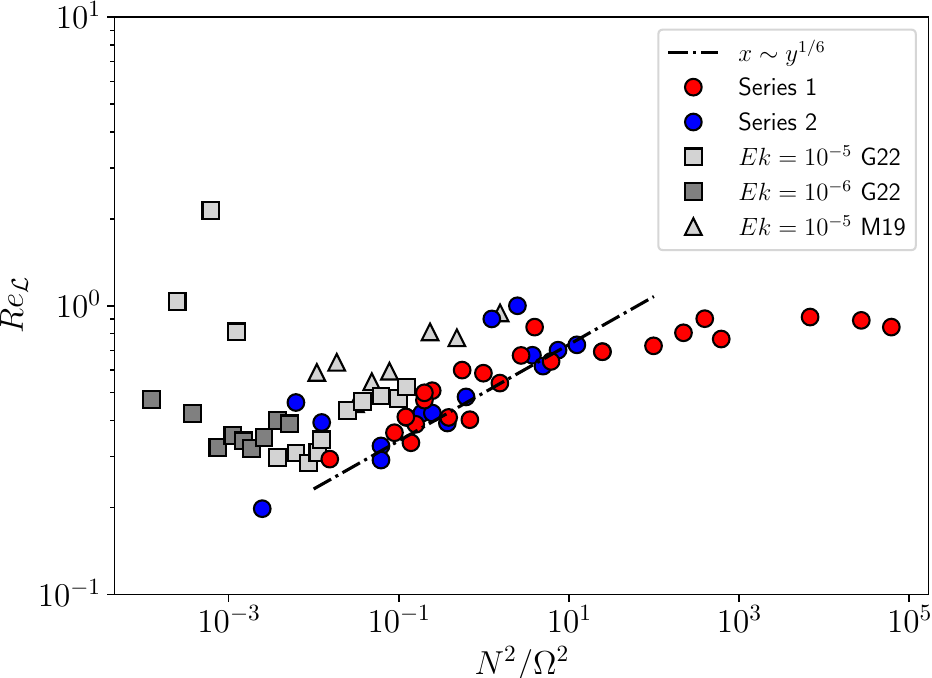}
	\caption{\revision{Variation of the local Reynolds number $\Rel$ as a function of $N^2/\Omega^2$. The Reynolds numbers is based on the mean poloidal velocity. The  theoretical scaling of \citet{Tassin2024}, $\Rel\sim\Ra_c^{1/6}$, is indicated by a dashed line. Data from \citet{Monville2019} (M19) and \citet{Guervilly2022} (G22) are included for smaller $\Ek$.}}
	\label{fig:Re_N2}
\end{figure}

\revision{The saturation of the flow amplitude is commonly assessed via the dependence of the Reynolds number on the control parameters. Here we measure the Reynolds number $\Rey$ as a dimensionless measure of the r.m.s. poloidal velocity $V_{p}$, 
\begin{equation}
    \Rey = \frac{V_{p} D}{\nu},
\end{equation}
where $V_{p}$ is calculated from the volume and time average of the poloidal  kinetic energy.}
\revision{The variation of $\Rey$ with the parameters can be compared with}
the theoretical scalings that have been proposed in non-rotating \citep[\eg][]{Brown2013,Yang2016,Hage2010,Tassin2024} and rotating fingering convection \citep{Sengupta2018} for regimes of weakly and strongly driven fingering convection, defined by the reduced density ratio \citep{Traxler2011b}:
\begin{equation}
    \smallr = \frac{\Rrhoi-1}{\Le-1}.
\end{equation}
The regime of weakly driven convection corresponds to $\smallr \approx1$ (\ie close to the linear onset for large Rayleigh numbers, $\Rrhoi \approx \Le$), whilst the regime of strongly-driven fingering convection, close to neutral stability ($\Rrhoi=1$), corresponds to \mbox{$0<\smallr\ll1$}. Note that here $\smallr=0.26$, so our simulations have moderate driving by that definition. 
Nevertheless, we can attempt to compare our numerical data to theoretical scalings proposed for $\smallr\ll1$.

A simple theoretical scaling can be derived by assuming that the timescale for thermal diffusion and advection are similar at the finger scale $\lscale$, meaning that the typical radial velocity of the fingers is of the order of $\kappa_t/\lscale$, \ie the local P\'eclet number is \revision{order unity} \citep{Radko2013}:
\begin{equation}
    \Rel=\Rey \lscale \sim \frac{1}{\Pran}.
    \label{eq:Rel1}
\end{equation}
However, for small Prandtl numbers, \citet{Brown2013} and \citet{Sengupta2018} argue that 
the saturation of the fingering instability is caused by a secondary shear instability of the fingers \citep{Radko2012} and propose an alternative scaling depending on $\smallr$. 
In particular, for the range of parameters that is closest to our study, $\smallr\ll \Pran < 1$ and $\Le\gg1$, \citet{Brown2013} found that
\begin{equation}
    \Rel \sim \frac{1}{\Pran^{1/2}}.
    \label{eq:Rel2}
\end{equation}

\revision{An alternative scaling was proposed by \citet{Tassin2024} for non-rotating fingering convection by considering that the buoyancy sources must balance the viscous dissipation in a steady state. The scaling assumes that the viscous dissipation occurs at the finger size and that the convective transport follows the 1/3-law between the compositional Nusselt number and $\Ra_c$ (see discussion in Section~\ref{sec:Nusselt}). 
At $\smallr\ll1$, this theoretical scaling is 
\begin{equation}
    \Rey  \sim \Ra_c^{2/3} |\Ra_t|^{-1/4} \Sc^{-1}.
    \label{ref:Re_Tassin}
\end{equation}
\citeauthor{Tassin2024} found a good agreement between their data, those of \citet{Yang2016} and this theoretical scaling. The scaling can be re-written for the local Reynolds number, assuming that $\lscale\propto |\Ra_t|^{-1/4}$,
\begin{equation}
    \Rel = \Rey \lscale \sim \Ra_c^{1/6} \Rrho^{-1/2} \Pran^{-1/2} \Sc^{-1/2}.
    \label{eq:Rel_Tassin}
\end{equation}
}

\revision{Figure~\ref{fig:Re_N2} shows the variation of the local Reynolds number, $\Rel=\Rey \lscale$, where $\lscale$ is measured from the spectral peak, as a function of $N^2/\Omega^2$ in our simulations. 
The data from \citet{Monville2019} and \citet{Guervilly2022} are also included.}
\revision{First we note that the local Reynolds numbers are close to or 
below one, meaning that, at the finger scale, the poloidal flow remains laminar.
The local Reynolds number remains fairly constant throughout.
The data cannot distinguish between the two scalings~\eqref{eq:Rel1} and \eqref{eq:Rel2} since the Prandtl number is fixed and not particularly small ($\Pran =0.3$).
Similarly, the density ratio is fixed, so we can only test the dependence of $\Rey$ on one of the Rayleigh numbers. 
The numerical data exhibit a weak increasing trend that appears consistent with the scaling~\eqref{eq:Rel_Tassin} derived by \citet{Tassin2024}, although the shallow dependence on $\Ra_c$ makes numerical confirmation challenging. 
This trend is also visible in Series 1 for $N^2/\Omega^2<100$, despite $\Ra_c$ being held fixed. For $N^2/\Omega^2>100$, $\Rel$ remains approximately constant, suggesting that rotational effects become dynamically negligible as expected at high $N^2/\Omega^2$ in Series 1, \ie as $\Ek\to 1$. This behaviour would not necessarily persist in Series 2 if $\Ra_c$ could be increased further. The critical Rayleigh number for the onset of fingering convection, $\Ra_c^{\mathrm{crit}}$, scales as $\Ra_c^{\mathrm{crit}}\propto \Ek^{-1}$ for $\Ek\ll1$ \citep{Monville2019}. It is therefore worth noting that the increase of the supercriticality $\Ra_c/\Ra_c^{\mathrm{crit}}\sim \Ra_c\Ek$ in Series 1 may partially account for the weak increase in the local Reynolds number for $N^2/\Omega^2<100$ in this case.}

\revision{In summary, we find that the local Reynolds numbers in our simulations remain nearly constant and close to unity. The scaling~\eqref{eq:Rel_Tassin} proposed by \citet{Tassin2024} for non-rotating convection, which predicts a weak dependence of $\Rel$ on $\Ra_c$, is consistent with our simulations, but a more definitive confirmation would require a wider exploration of the parameter space. Remarkably, comparison with the results of \citet{Monville2019} reveals good agreement between full-sphere and spherical-shell configurations, despite differences in geometry, boundary conditions, and stratification profiles.}

In the following, we will briefly describe the flows in the weakly- and rapidly-rotating regimes (\S\ref{sec:slow} and \ref{sec:rapid}) before presenting the intermediate regime in more detail (\S\ref{sec:intermediate}). 
Section~\ref{sec:zonal} presents the formation and saturation of the axisymmetric and azimuthal (\ie zonal) flows.
In Section~\ref{sec:Nusselt}, we discuss the transport properties of the fingering convection in the different regimes.

\subsection{Weak rotation}
\label{sec:slow} 

\subsubsection{Radial fingers}
\begin{figure}
	\centering 
   	\includegraphics[width=0.95\textwidth]{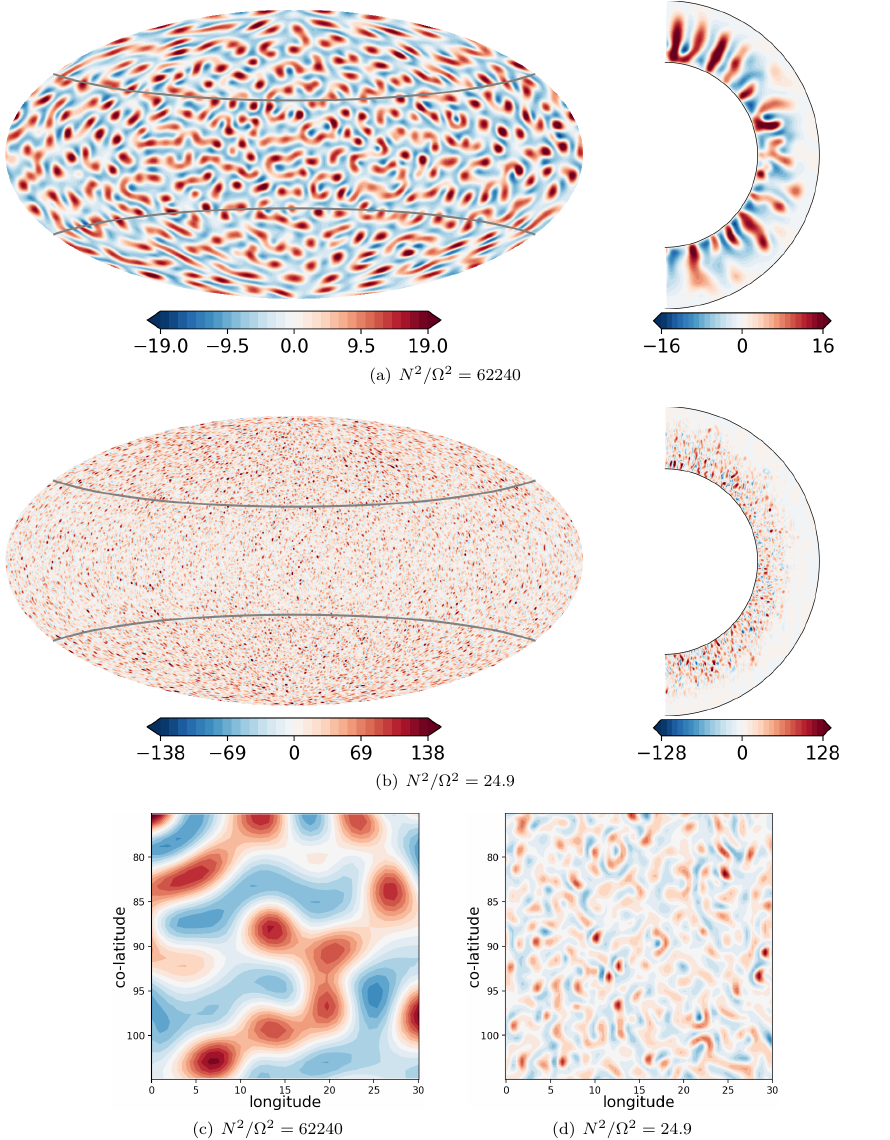}
	\caption{Snapshots of the radial velocity for simulations in the weakly-rotating regime: (a)-(b) Aitoff projection at $r=1.8$ (left) and meridional slice (right) for (a) $N^2/\Omega^2=62240$ in Series 1 ($\Ek=10^{-1}$, $\Ra_c=2\times10^7$) and (b) $N^2/\Omega^2=25$ in Series 2 ($\Ek=10^{-4}$, $\Ra_c=8\times10^9$). Plots (c) and (d) are close-ups of the spherical projections. In the Aitoff projections, the gray lines indicates the location of the tangent cylinder.}
	\label{fig:slow_shell}
\end{figure}

Figure~\ref{fig:slow_shell} shows the radial velocity projected on a spherical
surface at $r=1.8$ for simulations in Series 1 and 2 
representative of the weakly-rotating regime with $N^2/\Omega^2>1$. 
Since the thermal stratification is significantly stronger for our chosen case from Series 2
($|Ra_t|$ is 400 times larger), the finger size is visibly smaller. 
In either cases, no noticeable difference in the radial flows of the equatorial and polar regions is visible, suggesting that the dynamics of the fingers are weakly influenced by rotation.
The meridional slices of $u_r$ in Figure~\ref{fig:slow_shell} shows that, as expected from the radial profiles in Figure~\ref{fig:profile_ur}, the radial fingers are  preferentially located in the inner region where the density ratio is smallest. 
Similarly to the radial fingers described in \citet{Tassin2024} in the absence of rotation, the fat fingers at low $\Ra_c$ (Series 1) are tubular structures with a radial lengthscale that extend over more than half of the layer depth, whilst at larger $\Ra_c$ (Series 2), the thin fingers have undulating structures with shorter radial lengthscale.

\begin{figure}
	\centering 
        \includegraphics[width=\textwidth]{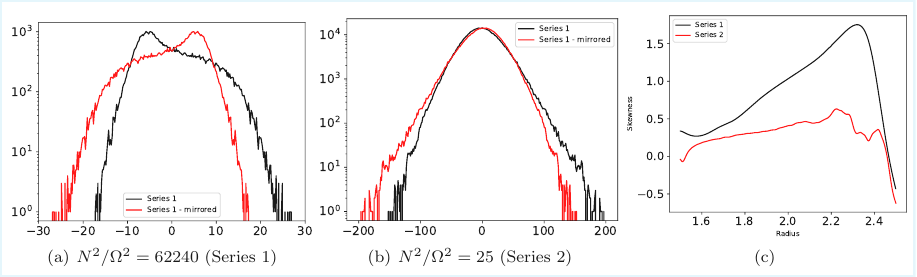}	
	\caption{(a) and (b) PDF of the radial velocity at $r=1.8$ for $N^2/\Omega^2>1$. The distribution is mirrored with respect to $x=0$ to highlight the asymmetry between positive and negative values. (c) Radial profile of the skewness $\gamma(r)$ of the radial velocity.}
	\label{fig:histo}
\end{figure}

The spherical projections of $u_r$ in Figure~\ref{fig:slow_shell} (especially in Series 1) show 
an asymmetry in the distribution of the upward and downward flows, 
with concentrated upflows amongst more dilute downflows. 
The net radial velocity must be zero to respect mass conservation, but the distribution can be skewed. The Probability Distribution Function (PDF) of the radial velocity at $r=1.8$ is shown in Figure~\ref{fig:histo}. The distribution is skewed for both cases from Series 1 and 2, with a long tail in the positive values. The skewness is very evident for Series 1, with a prominent shoulder corresponding an accumulation of weak negative values. 
We calculate the skewness of the distribution as 
\[ \gamma(r) = \dfrac{\bigl \langle u_r^3 \bigr \rangle_{\mathcal{S}(r)}}{\bigl \langle u_r^2 \bigr \rangle_{\mathcal{S}(r)}^{3/2}},\]
where the angular brackets correspond to an average over the spherical surface $\mathcal{S}(r)$.
Figure~\ref{fig:histo}(c) shows the radial profile of $\gamma(r)$ for these two cases. Note that a limited number of data snapshots were available for the case of Series 2, so the profile shows significant fluctuations at large radii. In either case, the radial velocity is positively skewed for most of the domain and the skewness increases with radius.
At radius greater than $r>2.3$, where the flow is weak, the skewness drops and becomes negative in the outer boundary layer.

\revision{The pattern of skewness for $r<2.3$ can be understood from the observations above, that the fingers are tubular (i.e.\ of constant width), and more pronounced and consistent for Series 1.
The finger width is broadly constant as the buoyant fluid rises because this is controlled by the Stern scale.
In the spherical geometry, the filling factor of the regions of upwelling fluid therefore decreases with radius;
and from conversation of mass, the outward radial velocity in the upwellings increasingly dominates over the inward velocity elsewhere,
leading to the positive skewness.
The stronger skewness observed for Series~1 in Figure~\ref{fig:histo}(c) is linked to the heavier positive tail noted in Figure~\ref{fig:histo}(a), and the more consistent extension of the fingers throughout the shell.}

\subsubsection{Finger clusters and toroidal gyres}
\label{sec:cluster}

\begin{figure}
	\centering 
	\includegraphics[width=\textwidth]{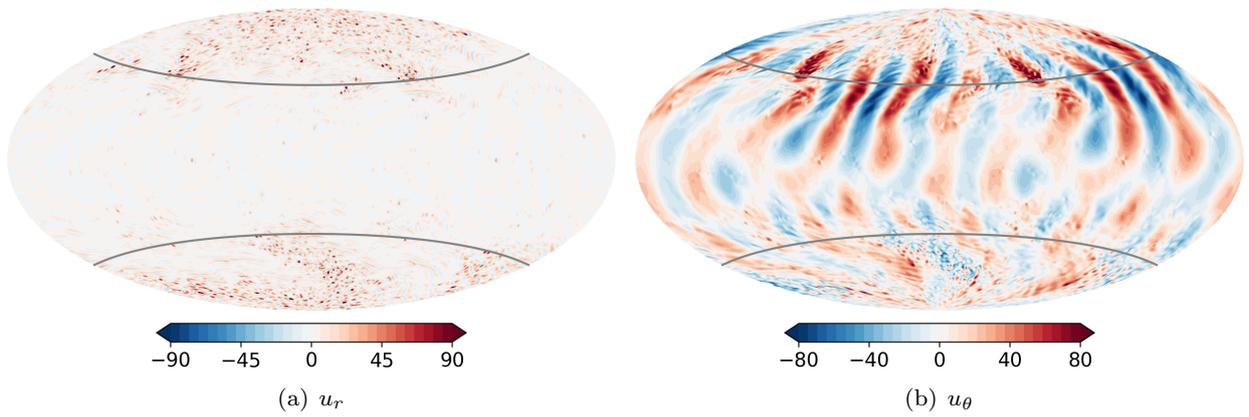}
	\caption{Aitoff projection at $r=2.1$ of a snapshot of the radial and latitudinal velocities for $N^2/\Omega^2=12.4$ in Series 2.}
	\label{fig:gyres_shell}
\end{figure}

\begin{figure}
	\centering 
	\includegraphics[width=0.8\textwidth]{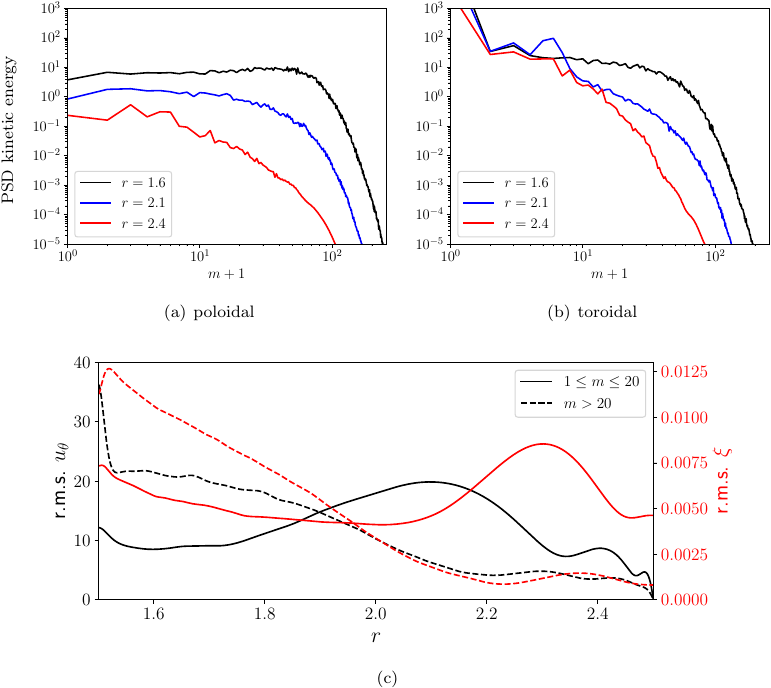}
	\caption{(a)-(b) Spectra of the poloidal and toroidal kinetic energy as a function of the azimuthal order $m$ at different radii for $N^2/\Omega^2=12.4$ in Series 2.
    (c) Radial profiles of the filtered r.m.s. latitudinal velocity and composition perturbation. The composition perturbation is normalised by $C_s$ at $r=r_i$.}
	\label{fig:spec_gyres}
\end{figure}

In some cases with $N^2/\Omega^2>1$, the dynamics change in the upper half of the domain ($r\gtrsim 2$) with the localisation of radial fingers into large-scale patches (or clusters) at high latitudes. This is illustrated in Figure~\ref{fig:gyres_shell} showing the spherical projection of $u_{r}$ and $u_{\theta}$ at $r=2.1$ for $N^2/\Omega^2=12.4$ in Series 2. Fingering convection in the lower half of the domain is more vigorous and there is no evidence of clustering, similarly to the dynamics observed in Figure~\ref{fig:slow_shell}(b).
The patches in the upper domain are much larger scale than the transverse finger size and are accompanied by toroidal gyres.
Figure~\ref{fig:spec_gyres} shows the poloidal and toroidal kinetic energy spectra as a function of the azimuthal order $m$ at different radii for the case of Figure~\ref{fig:gyres_shell}. In the lower domain, the poloidal and toroidal flows involve a wide range of azimuthal wavenumbers from $m=1$ to $m\approx50$, whilst in the upper domain, the flow is dominated by smaller azimuthal wavenumbers ($m\lesssim 20$) and is prominently toroidal. The zonal flow (toroidal $m=0$ mode) dominates at all radii and will be discussed separately in section~\ref{sec:zonal}. 

The clusters are associated with the presence of large-scale density anomalies that emerge in the upper domain.
The fingers tend to cluster in the heavier anomalies, and are nearly absent in the lighter anomalies. Large-scale upwelling and downwelling flows are associated with these density anomalies, but they are very weak compared with the fingers in the lower domain as indicated by the spectra of Figure~\ref{fig:spec_gyres}; they are also not visible in the radial profiles of the r.m.s. velocity in figure~\ref{fig:profile_ur} for $N^2/\Omega^2>1$. The large-scale anomalies are more evident in the composition perturbation, so we use the composition to understand the spatial relation between anomalies and toroidal gyres.
Large-scale temperature perturbations also exist, which counteract the composition perturbations; however, their amplitude is smaller, resulting in the composition perturbations dominating the density anomalies. 
Figure~\ref{fig:spec_gyres}(c) shows the radial profiles of the r.m.s. $u_{\theta}$ and $\Cpert$ separated into large azimuthal scales ($1\leq m\leq 20$) and smaller scales ($m>20$). The large-scale latitudinal velocity dominates in the upper domain and the r.m.s. profile has two local peaks (located at $r\approx 2.1$ and $r\approx 2.4$) that surround a peak of the large-scale composition perturbation. 

\begin{figure}
	\centering 
	\includegraphics[width=\textwidth]{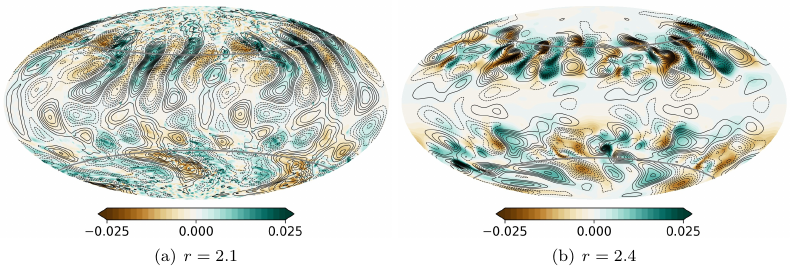}
	\caption{Isocontours of non-axisymmetric $\Cpert$ (colour) and $U_t$ (black lines) in an Aitoff projection at $r=2.1$ and $r=2.4$ for $N^2/\Omega^2=12.4$ in Series 2. 
    The composition is normalised by $C_s$ at $r=r_i$.
    For $U_t$, the solid (dashed) lines indicates positive (negative) contours. Anti-clockwise gyres have positive $U_t$.}
	\label{fig:surf_tor_gyres}
\end{figure}

Since the flow is dominated by the toroidal component at large radii, we plot in Figure~\ref{fig:surf_tor_gyres} isocontours of the non-axisymmetric toroidal scalar $U_t$ and composition perturbation at radius $r=2.1$ and $r=2.4$. The location of the toroidal gyres and composition patches is clearly correlated. 
At $r=2.1$, anti-clockwise gyres ($U_t>0$) are associated with lighter patches ($\xi>0$) in the northern hemisphere, and vice-versa in the southern hemisphere. This hemispherical dependence indicates the importance of the Coriolis force in the formation of these toroidal flows. 
The composition patches extend through the upper domain and keep their coherence. 
The toroidal gyres change sign of rotation at $r=2.4$, \ie clockwise gyres ($U_t<0$) are now associated with lighter patches.
These observations suggest that the slow, large-scale upflows of lighter fluid draw convergent horizontal flows at the lower roots around $r\approx2.1$ and divergent horizontal flows at their top ($r\approx2.4$) (poloidal $u_{\theta}$ and $u_{\phi}$ not shown here); and vice-versa for downflows of heavier fluid. 
These convergent/divergent flows are deflected by the Coriolis force into anti-clockwise/clockwise toroidal gyres in the northern hemisphere (and vice-versa in the southern hemisphere). 

The finger clusters occur in a limited range of the parameter space indicated by open circles in Figure~\ref{fig:Ek_vs_Ra}. They are not visible in Series 1, which suggests that sufficiently low Ekman number might be required. To test their sensitivity to changes in $N^2/\Omega^2$, we ran a number of simulations at intermediate values of $\Ra_c$ indicated in Figure~\ref{fig:Ek_vs_Ra}. 
The clusters are present for $N^2/\Omega^2\approx 100$, but disappear for $N^2/\Omega^2\approx 1000$, suggesting that they occur in the limited range $N^2/\Omega^2\approx 1-100$ for $\Ek\lesssim 6\times 10^{-4}$.

Finger clustering was previously observed in Cartesian box simulations at high $\Pran$ and low density ratio \citep{Paparella2012}, but this is likely through a different mechanism than observed here. Indeed, in these box simulations, the  clustering is defined as the formation of large-scale buoyant structures and is thought to occur via entrainment as the local Reynolds number might exceed unity. Unlike in our simulations, this occurs in the absence of rotation. Here the clustering corresponds to a spatial localisation of fingering convection. It is linked to the development of large-scale density anomalies: within the heavier anomalies, the local compositional Rayleigh number, estimated from the local radial compositional gradient, is up to four greater than in the lighter anomalies. Consequently, fingering convection is preferentially promoted in the heavier anomalies. 
\revision{This feedback is suggestive of the presence of large-scale instabilities \citep[\eg][]{Traxler2011,Radko2013}. The occurrence of large-scale buoyant patches in the presence of latitudinal gradients of temperature and composition (arising from rotationally induced latitudinal inhomogeneities in the mixing, see Section~\ref{sec:Nusselt}) is reminiscent of lateral intrusions, which manifest as laterally interleaving layers \citep{Holyer1983,Ruddick2003,Medrano2014}. Moreover, lateral intrusions are known to develop at large density ratios ($\Rrho>\Le$), outside the parameter regime unstable to the primary fingering instability, consistent with the observation that the large-scale patches occurs in the upper part of the layer in our simulations.
Although intrusions typically form quasi two-dimensional, tilted horizontal layers, it is plausible that spherical geometry and Coriolis effects disrupt these layers into patches of finite azimuthal extent. The possible formation of intrusions in our simulations warrants further investigation, which we plan to pursue in future work, particularly in simplified configurations (\eg in the strongly-stratified regime $\Rrho>\Le$, as in \citet{Medrano2014}).}

\subsection{Rapid rotation}
\label{sec:rapid} 

\subsubsection{Columnar fingers}
\begin{figure}
	\centering 
	\includegraphics[width=0.8\textwidth]{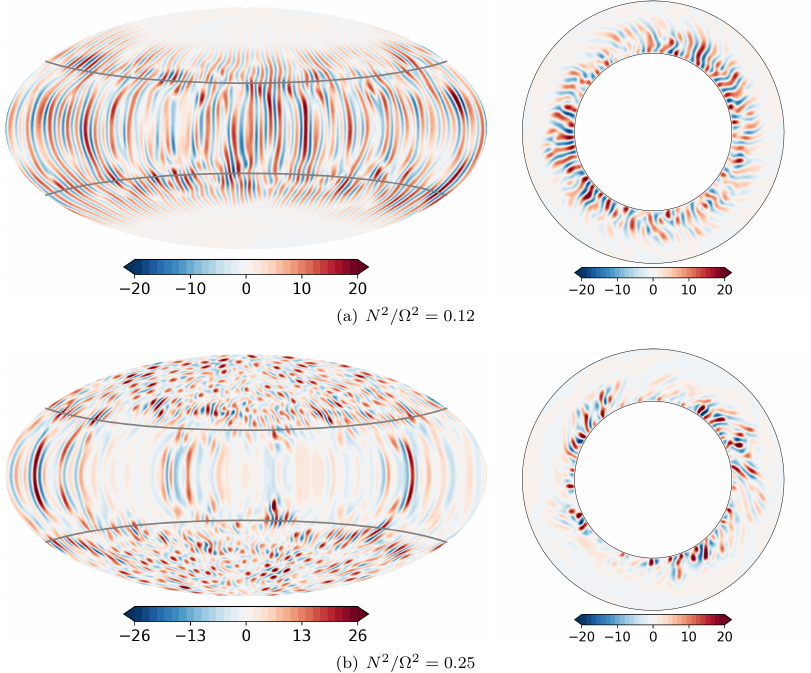}
	\caption{Snapshot of the radial velocity for simulations in the rapidly-rotating regime in Series 2 ($\Ek=10^{-4}$): Aitoff projection at $r=1.8$ (left) and equatorial slice (right).}
	\label{fig:fast_shell}
\end{figure}

Flows in the rapid-rotation regime were described in \citet{Guervilly2022} in the same configuration
and using a larger dataset including lower Ekman numbers ($\Ek\in [10^{-6},10^{-4}]$).
The main features are summarised in this short subsection for completeness.

Figure~\ref{fig:fast_shell} shows spherical projections and equatorial slices of the radial velocity 
for two cases of Series 2 with $N^2/\Omega^2<1$. For the smallest stratification,
columnar fingers occur mainly outside of the tangent cylinder, impinging slightly
inside the tangent cylinder. Larger $N^2/\Omega^2$ leads to the development
of columnar fingers in the polar regions. 
The radial flows gradually become stronger in the polar regions than in the equatorial regions as $N^2/\Omega^2$ increases, whilst
fingering convection become localised in the equatorial region.
Zonal flows develop in the equatorial region (see section~\S\ref{sec:zonal}), 
leading to a visible shear and disruption
of fingering convection in that region, which might explain why radial flows are relatively more vigorous in the polar regions.
\revision{The localised fingering structures in the equatorial region drift azimuthally as they are advected by the zonal flow.}

\subsubsection{Equatorially antisymmetric mode}
\label{sec:EAS}

\begin{figure}
	\centering 
	\includegraphics[width=\textwidth]{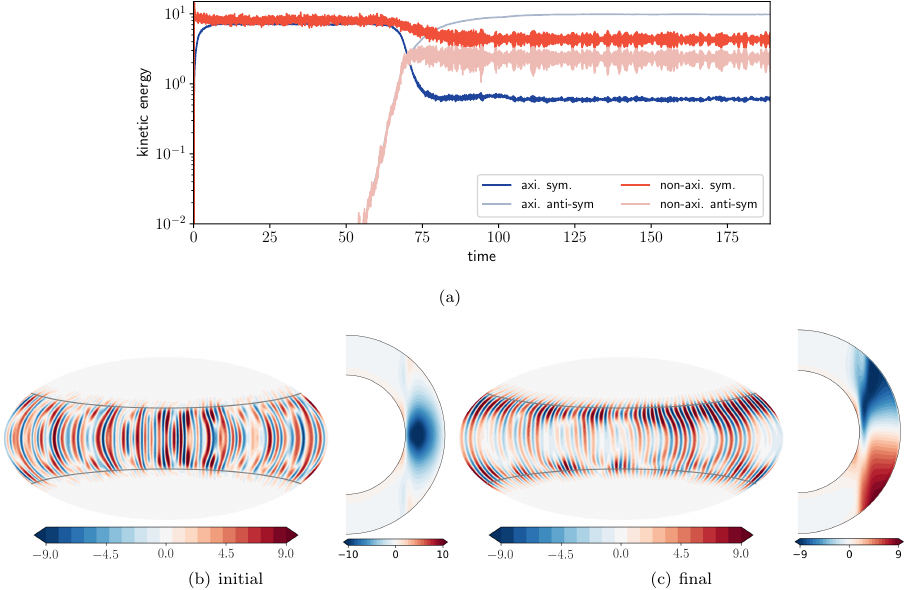}
	\caption{(a) Time series of the volumetric kinetic energy for $N^2/\Omega^2=0.06$ and $\Ek=10^{-4}$. Sym/anti-sym refers to the equatorial symmetry. 
    (b)-(c) snapshots of the radial velocity at $r=1.8$ and meridional slice of the zonal velocity 
    in the initial steady state, shortly after the beginning of the simulation and in the final state.}
	\label{fig:Rac5e7}
\end{figure}

We observe an interesting behaviour (not reported in \cite{Guervilly2022}) from an extended running time for the simulation with $N^2/\Omega^2=0.06$ (which is shared by Series 1 and 2) 
as shown in the time series of Figure~\ref{fig:Rac5e7}.
The flow is columnar and located  outside the tangent cylinder.
In the initial steady state (Figure~\ref{fig:Rac5e7}(b)), the flow is symmetric with respect to the equatorial plane (equatorially symmetric -- ES) 
with a dominant non-axisymmetric component.
Then, after over 60 viscous timescales, the flow gradually transitions to a different steady state. 
The transition corresponds to a change in the equatorial symmetry of the flow.
After the transition (Figure~\ref{fig:Rac5e7}(c)), the zonal velocity dominates and is now largely 
equatorially antisymmetric (EAS). 
Whilst the non-axisymmetric flow remains mainly ES after the transition, its amplitude is divided by a factor two (so is then only a factor of two larger than the non-axisymmetric EAS component).
Note that in all the plots showing global averages ($\Rey$, $\Rey_0$ and $\Nu_c$ below), we use the values averaged before the transition for simplicity. The output quantities calculated after the transition to the EAS mode are given in Table~\ref{tab:Series2}.

The transition to the EAS mode
was previously observed in the simulations of fingering convection in a full sphere geometry of \cite{Monville2019}. 
\citeauthor{Monville2019} reported that this transition occurs for Rayleigh numbers 
above the linear onset for the EAS axisymmetric mode, which suggests that this mode grows linearly. 
This mode corresponds to hemispherical convection with EAS axisymmetric composition and temperature perturbations,
where the northern hemisphere is enriched in light elements and colder than the southern hemisphere.
This is the only instance where the EAS axisymmetric mode becomes dominant in our simulations,
even after running simulations for multiple tens of viscous timescales.
For smaller stratification, the mode decays; for larger stratification, the mode saturates at a smaller amplitude
than the ES axisymmetric mode.  
This simulation shows that the EAS axisymmetric mode can occur in spherical shells, even for relatively large aspect ratio
(here $r_i/r_o = 0.6$). As argued in \cite{Monville2019}, simulations spanning many viscous times are required to observe this transition.

\subsection{Intermediate regime}
\label{sec:intermediate}

\subsubsection{Banded structures}
\begin{figure}
	\centering 
    \includegraphics[width=0.8\textwidth]{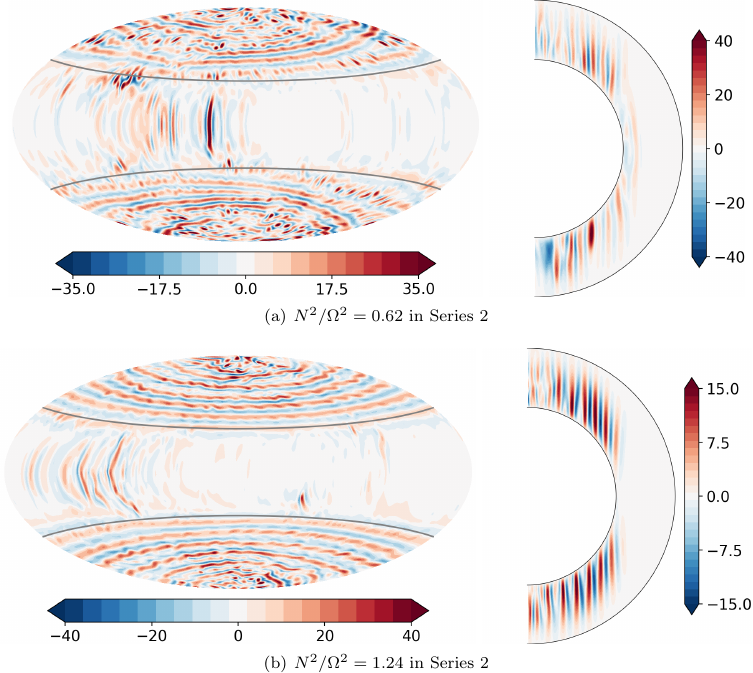}
	\caption{Snapshots of the radial velocity for simulations in the intermediate regime: (left) Aitoff projections at $r=1.8$; (right) meridional slice at a given longitude. 
    For (b), the meridional slice shows the axisymmetric component.} 	
    \label{fig:transition_shell}
\end{figure}

\begin{figure}
	\centering 
    	\includegraphics[width=0.8\textwidth]{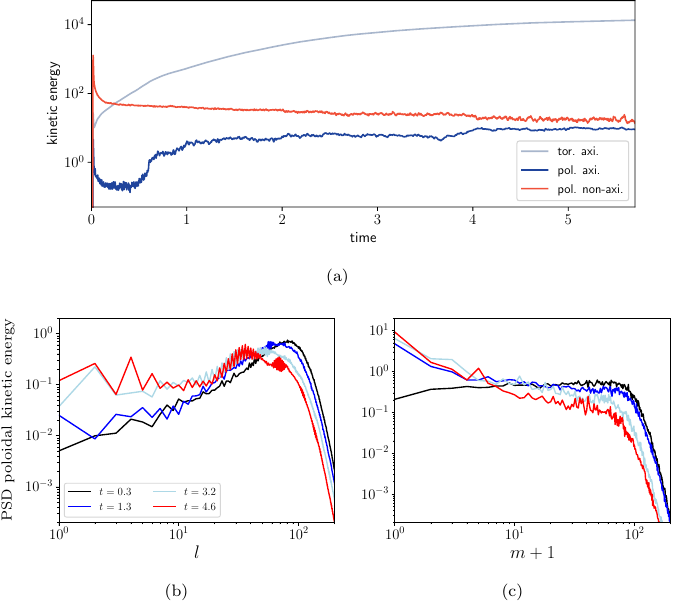}
	\caption{(a) Times series of the volumetric kinetic energy for $N^2/\Omega^2=1.24$ of Series 2 decomposed into 
 poloidal/toroidal components and axisymmetric/non-axisymmetric components.
    (b)-(c) Spectra of the poloidal kinetic energy as a function of the spherical harmonics degree $l$ and order $m$ at different times of the simulation. The spectra are averaged in radius excluding the boundary layers.}
	\label{fig:spectra_bands}
\end{figure}

We now explore the intermediate regime where $N^2/\Omega^2\approx 1$. Over timescales of the order of a viscous 
timescale, we observe the formation
and saturation of bands of poloidal flows as shown in the spherical projection of $u_r$ in
Figure~\ref{fig:transition_shell}. 
These bands are persistent and only form inside the tangent cylinder. 
While in some cases the bands
are spiraling (corresponding predominantly to an azimuthal order $m=1$, Figure~\ref{fig:transition_shell}(a)), 
in other cases the bands are predominantly axisymmetric (corresponding to $m=0$, Figure~\ref{fig:transition_shell}(b)).
The bands have a narrow latitudinal extent compared with their azimuthal extent.
The meridional slices of Figure~\ref{fig:transition_shell} show the axial extent of the bands. The bands are essentially concentric cylindrical structures aligned with the rotation axis.
The radial flows in the equatorial regions are comparatively weak, which might be due to the presence of strong zonal flows that can disrupt convection in this region.
Indeed, the amplitude of the zonal flow is maximal in the intermediate regime, exceeding the amplitude of the radial flows by one order of magnitude (see section~\S\ref{sec:zonal}). 

Figure~\ref{fig:spectra_bands} shows the times series of the kinetic energy and 
the power spectra of the poloidal kinetic energy at different 
times of the simulation at $N^2/\Omega^2=1.24$ of Series 2. The fingers initially develop at a size 
close to the Stern scale (around $l\approx m \approx 80$), but the energy then builds up 
at small $m$, with a dominant $m=0$, and around about $l\approx 40$.
Meanwhile the energy at the Stern scale gradually decreases before reaching saturation at a lower
energy level than the bands. The fingers might be losing energy to the benefit of the 
bands, either due to a decrease of the mean composition gradient in the bulk or due to non-linear energy transfers, 
or because of the progressive increase of the zonal flow, which finally reaches saturation after $t>5$.
At saturation ($t=4.6$), the bands carry a similar amount of kinetic energy to the initial fingers at the Stern scale (at $t=0.3$). Their associated local Reynolds number is thus greater (as their latitudinal scale are larger), but remains below unity. 

\begin{figure}
	\centering 
	\includegraphics[width=0.6\textwidth]{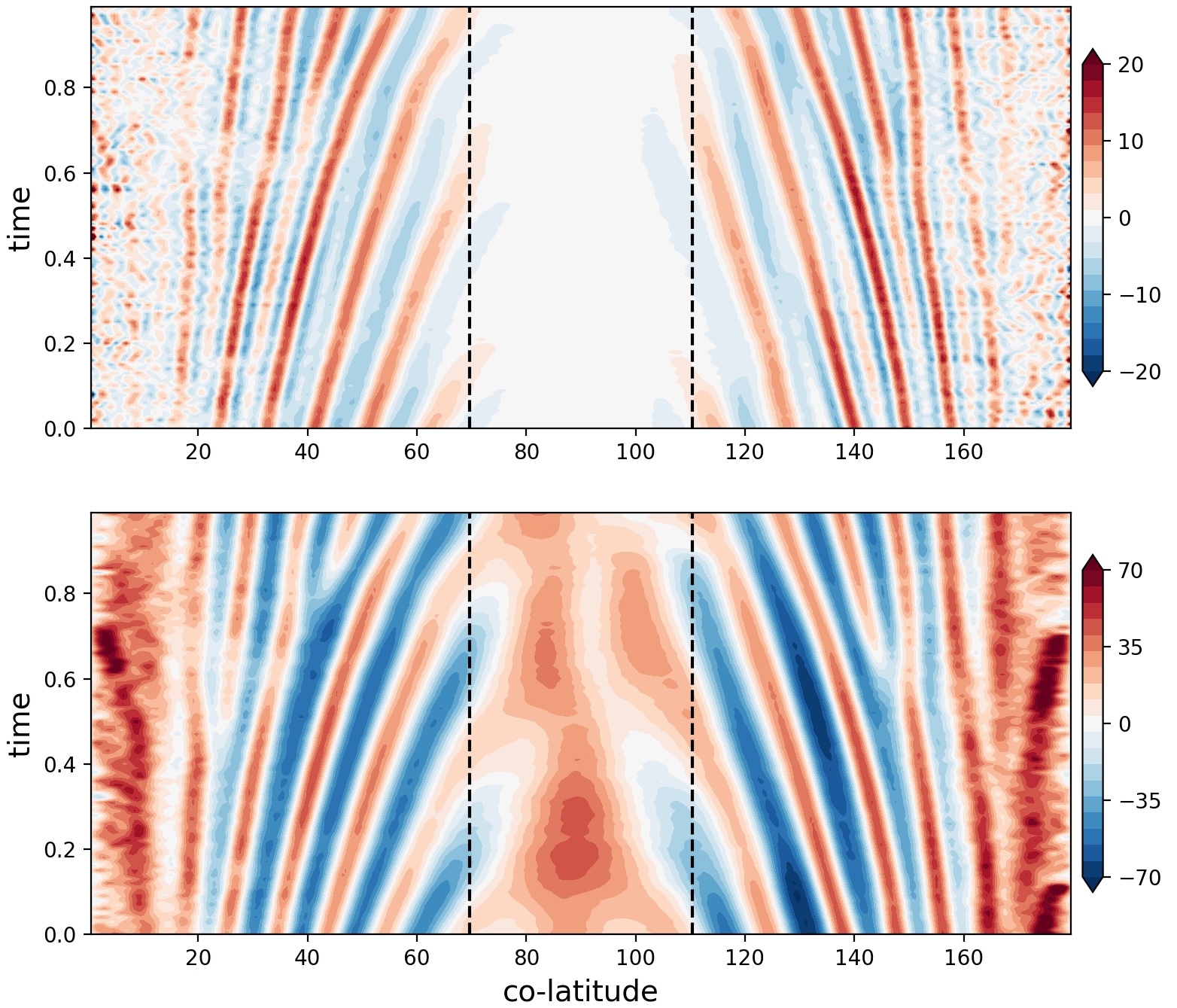}	
	\caption{Space-time diagram of the axisymmetric radial (top) and azimuthal (bottom) velocities at $r=1.6$
    for $N^2/\Omega^2=1.24$ of Series 2.  
    The spherical harmonics degree $l=1$ has been removed from the azimuthal component
    to filter out the dominant equatorial prograde zonal flow.
    The dashed line corresponds to the intersection with the tangent cylinder for this radius.}
    \label{fig:transition_drift}
\end{figure}

\revision{Figure~\ref{fig:transition_drift} shows a space-time diagram of the axisymmetric radial velocity at a fixed radius close to the inner boundary ($r=1.6$), where the horizontal axis corresponds to co-latitude. 
This representation allows us to track the temporal evolution of the axisymmetric bands in the saturated regime and reveals their latitudinal drift.}
The time period corresponds to approximately one viscous timescale (one unit of dimensionless time).
For latitudes smaller than $\pm70^{\circ}$, the bands slowly drift latitudinally over time towards the equatorial region, 
eventually disappearing on the tangent cylinder. New drifting bands nucleate over time around latitudes of $\pm70^{\circ}$.
At higher latitudes, weak meandering bands can be seen and appear to be drifting towards the poles.  
\revision{The latitudinal drift is not attributable to meridional circulation, as the meridional velocity is very weak within the tangent cylinder and would not produce the observed drift direction, making this mechanism unlikely. While advection by the zonal flow could account for the drift of the spiraling structures, it does not explain the latitudinal drift observed in the axisymmetric bands in Figure~\ref{fig:transition_drift}.}

Azimuthal velocities are associated with the presence of the bands of poloidal flows. 
Similarly, they show a spiraling or axisymmetric pattern.
They can be seen inside the tangent cylinder in Figure~\ref{fig:zonal}(c) in \S\ref{sec:zonal}
with a smaller amplitude than the equatorial prograde zonal.
To observe their temporal evolution, we filter out the spherical harmonics degree $l=1$ in the post-processed data, 
which is the main component of the equatorial zonal flow. The space-time diagram of the filtered 
axisymmetric azimuthal velocity is shown in Figure~\ref{fig:transition_drift} and reveals that the latitudinal drift is also seen in the banded azimuthal flow.

The bands have a narrow range of existence, only appearing for parameters around $N^2/\Omega^2\approx 1$ (see Figure~\ref{fig:Ek_vs_Ra}). 
Varying the buoyancy frequency by more than one order of magnitude above or below this threshold value
leads to a transition to radial or columnar fingers respectively.
The bands are robust features in the sense that they occur for $N^2/\Omega^2\approx1$ 
irrespective of $\Ek$ or $Ra_c$ (\ie they appear in both Series 1 and 2 and in additional simulations shown in Figure~\ref{fig:Ek_vs_Ra}), at least for the density ratio used here. 

\subsubsection{Origin of the bands}

\begin{figure}
	\centering 
	\includegraphics[width=0.9\textwidth]{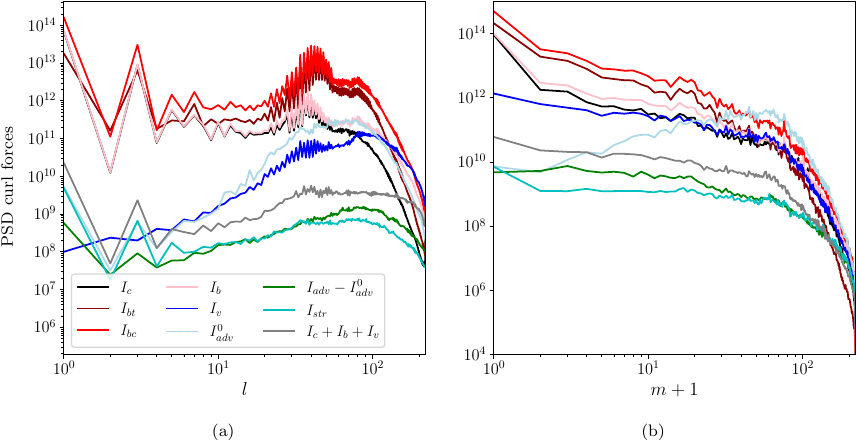}
	\caption{Spectral representation of each terms in the equation for $\phi$-component of the vorticity (equation~\eqref{eq:wphi}) for the case 
    $N^2/\Omega^2=1.24$ of Series 2. The spectra are averaged in radius between 1.6 and 2.
    $I_c$ corresponds to the Coriolis term; $I_{bt}$ and $I_{bc}$ to the thermal and compositional buoyancy terms; $I_v$ to the viscous term; $I_{adv}^0$ and $I_{adv}-I_{adv}^0$ to the advection by the zonal and non-zonal velocity; $I_{str}$ to vortex stretching. The total buoyancy $I_b=I_{bt}+I_{bc}$ and the residual $I_c+I_b+I_v$ are 
    calculated in the physical space before calculating the spectral decomposition.}
	\label{fig:FB_spectra}
\end{figure}

We conclude this section by studying the origin of the bands. In particular, we want to understand
the role of the non-linear interactions in the formation of the bands. 
First, we compare the relative importance of the forces in the Navier-Stokes equation at different scales. 
This is most easily done by comparing the sources of the vorticity $\vor=\nabla \times \vel$. 
Since the bands correspond essentially to axisymmetric or $m=1$ poloidal flows, 
we analyse the relative strengths of the terms in the equation for the $\phi$-component of the vorticity:
\begin{eqnarray}
    \pd{\vorp}{t} + \left[ (\vel \cdot \nabla ) \vor \right]_{\phi} - \left[ (\vor \cdot \nabla ) \vel \right]_{\phi}
    - \frac{2}{\Ek} \pd{u_{\phi}}{z} = \left[ \nabla^2 \vor \right]_{\phi} 
    - \frac{1}{r_o} \pl \pd{\Tpert}{\theta} + \pd{\Cpert}{\theta} \pr.
    \label{eq:wphi}
\end{eqnarray}

For a given snapshot, we calculate the power spectra of each term as a function of the spherical harmonics degree $l$ or order $m$. The goal here is to determine whether the force balance differs 
for the bands and the fingers.
Figure~\ref{fig:FB_spectra} shows the power spectra for the case of Figure~\ref{fig:spectra_bands}.
The spectra are averaged in radius between $1.6$ and $2$, where the bands have the strongest amplitude, and calculated at a time where the system is in a statistically steady-state. 
For spherical harmonics degrees and orders smaller than $l\lesssim 100$ and $m\lesssim 30$, the dominant vorticity sources are the thermal and compositional
buoyancy terms (last term on the right-hand side (RHS) of equation~\eqref{eq:wphi}).
They partially cancel each other as shown by the lower value of the total buoyancy for these scales. The thermal diffusion wipes out smaller scales and the thermal buoyancy
drops out more rapidly than the compositional buoyancy in the spectral tail. 
At the scale of the bands ($m=0,1$ and $l\approx40$), the main balance for the generation of the
azimuthal vorticity is between the buoyancy term and the Coriolis term (last term on the left-hand side (LHS) of \eqref{eq:wphi}). 
At the finger scale ($l\approx80$), the viscous term (first term on the RHS) also enters this balance.
The non-linear inertial terms (2nd and 3rd terms in the LHS) are significantly smaller than buoyancy, viscous and Coriolis terms
at all scales, except for the advection by the zonal velocity which is dominant at the finger scale.
Using snapshots at earlier times during the growth of the bands gives the same picture, suggesting again that inertia is unimportant for the formation of the bands. 

To confirm this, we performed a linear simulation for $N^2/\Omega^2=1.24$ of Series 2, where
\revisiontwo{the non-linear terms $(\vel \cdot \vn )\vel$, $\vel \cdot \vn \Tpert$ and $\vel \cdot \vn \Cpert$ are switched off in equations~\eqref{eq:NS}-\eqref{eq:Cpert} in XSHELLS}
\revision{(\ie calculating the growth or decay of perturbations from the static state)}: 
the mode $m=0$ grows and forms banded structures inside the tangent cylinder similar to those described 
in the non-linear calculations. 
We therefore conclude that the bands corresponds to a linear mode of fingering convection, which is preferred when $N^2/\Omega^2\approx1$ for this density ratio.  
\revision{In the linear calculation, the bands do not drift in the latitudinal direction, suggesting that the drift observed in the full non-linear simulations arises from a non-linear mechanism.}

\subsection{Zonal Flow}
\label{sec:zonal}

\subsubsection{Formation of the zonal flow}
Fingering convection produces persistent zonal flows in the presence of rotation. They are equatorially symmetric (except for the case $N^2/\Omega^2=0.06$ reported in \S\ref{sec:EAS}) and their speed can exceed the amplitude of the radial velocities. 
\citet{Guervilly2022} showed that the zonal flow is produced by a linear process and is in a thermo-compositional wind balance, where the latitudinal variations of the axisymmetric density perturbations are balanced by axial variations of the zonal velocity. 
The thermo-compositional wind balance is derived from the equation for $\omega_{\phi}$ (equation~\ref{eq:wphi}) in an axisymmetric, steady state and
neglecting the non-linear and viscous terms,
\begin{eqnarray}
    \frac{2}{\Ek} \pd{\overline{u_{\phi}}}{z} \approx  
    \frac{1}{r_o} \pl \pd{\overline{\Tpert}}{\theta} + \pd{\overline{\Cpert}}{\theta} \pr,
    \label{eq:TCwind}
\end{eqnarray}
where the overline denotes an azimuthal average. 
For rapidly-rotating fingering convection, the latitudinal difference in the mixing of composition and temperature is a source of baroclinicity, driving a meridional circulation that is deflected into $z$-dependent zonal flows by the Coriolis force.
Composition and temperature gradients oppose each other, but the slower diffusion of the composition leads to sharper composition gradients, which dictate the direction of the meridional circulation and zonal flows. 

\begin{figure}
	\centering 
	\includegraphics[width=\textwidth]{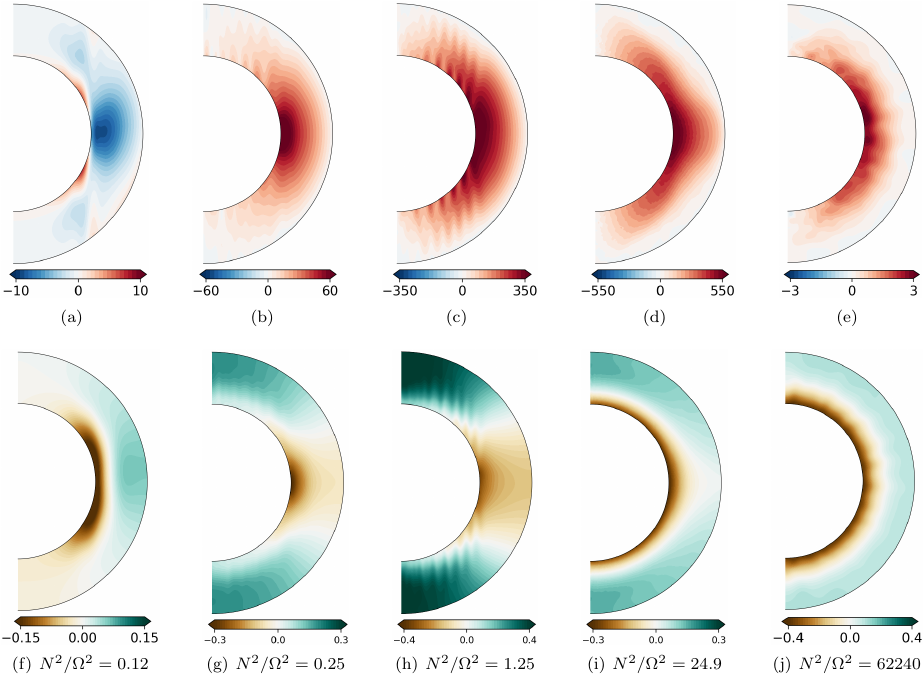}	
	\caption{Meridional slices of the zonal velocity (top row) and the axisymmetric composition perturbation (bottom row) for simulations in Series 2 ($N^2/\Omega^2\in[0.12,25]$) and Series 1 ($N^2/\Omega^2=62240$). 
    The composition is normalised by the basic state composition $C_s$ at $r=r_i$.}
	\label{fig:zonal}
\end{figure}

Figure~\ref{fig:zonal} shows meridional slices of the zonal velocity and the axisymmetric composition perturbation for increasing $N^2/\Omega^2$. 
For the smallest $N^2/\Omega^2$, the zonal flow is retrograde in the equatorial region and is associated with a compositional mixing that solely takes place outside the tangent cylinder, as the polar regions are not convecting (Figure~\ref{fig:fast_shell}(a)). As fingering convection onsets in the polar regions at larger $N^2/\Omega^2$ (Figure~\ref{fig:fast_shell}(b)), the mixing becomes more efficient at high than low latitudes and the latitudinal compositional gradient is reversed. This leads to a reversal in the direction of the zonal flow, which is strongest outside the tangent cylinder. Further increase of $N^2/\Omega^2$ up to approximately $10$ leads to an amplification of the latitudinal dependence in the mixing and an intensification of the prograde zonal flow.    
As $N^2/\Omega^2$ is increased further in the weak-rotation regime, the mixing becomes increasingly homogeneous with latitude, leading a gradual weakening of the zonal flow. At our largest stratification ($N^2/\Omega^2=62240$), the zonal flow is significantly weaker than the radial flows; we expect that it would eventually completely disappear as $N^2/\Omega^2\to\infty$.

\begin{figure}
	\centering 
        \includegraphics[width=0.9\textwidth]{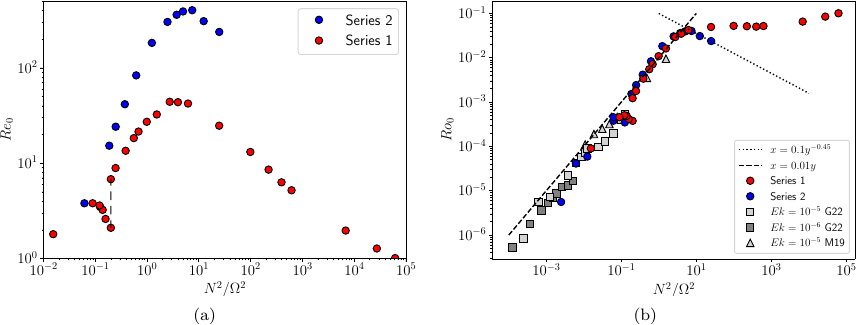}
	\caption{Variation of (a) the zonal Reynolds number and (b) the zonal Rossby number as a function of $N^2/\Omega^2$. The two solutions of the bistable case at $N^2/\Omega^2=0.2$ are linked by a dashed line in (a). The dashed line in (b) corresponds to $\Ro_0 = \Rey_0 \Ek = 10^{-2} N^2/\Omega^2$,
    \revision{while the dotted line corresponds to $\Ro_0=0.1(N^2/\Omega^2)^{-0.45}$, which is discussed in the conclusion section. Data from \citet{Monville2019} (M19) and \citet{Guervilly2022} (G22) are included for smaller $\Ek$ in (b). In \citet{Monville2019}, the Rossby number is calculated from the total kinetic energy, while in \citet{Guervilly2022} and in the present study, the Rossby number is based on the mean zonal velocity.}}
	\label{fig:Re0}
\end{figure}

\subsubsection{Scaling of the zonal velocity}
Figure~\ref{fig:Re0}(a) shows the variation of the Reynolds number based on the zonal flow amplitude, $\Rey_{0}=V_0 D/\nu$, where $V_0$ is the mean zonal velocity calculated from the volume and time average of the axisymmetric toroidal kinetic energy. 
$\Rey_0$ increases until $N^2/\Omega^2\approx10$ for both Series, and then decreases for larger values. 
At its maximum, $\Rey_0$ reaches a value approximately $10-30$ times larger than $\Rey$. This suggests that large values of the zonal flow can be reached at low Ekman numbers. 

Around $N^2/\Omega^2=0.2$, a discontinuity is visible in the variation of $\Rey_{0}$. This corresponds to the onset of fingering convection inside the tangent cylinder, which is accompanied by a reversal of the direction of the zonal flow, from retrograde (as in Figure~\ref{fig:zonal}(a)) to prograde (Figure~\ref{fig:zonal}(b)). 
The case $N^2/\Omega^2=0.2$ of Series 1 exhibits a transition from the initial (weaker) retrograde state to the final (stronger) prograde state after a long integration time of approximately $35$ viscous timescales.

In the rapidly-rotating regime, \citet{Guervilly2022} obtain an approximate scaling for the zonal flow amplitude by assuming a compositional wind balance, where only the latitudinal gradients of the composition balance with the axial gradient of the zonal velocity on large scales. Since the diffusion of the composition perturbations is slow compared with its advection, the scaling then assumes that the composition perturbations are of the same order of magnitude as the variation of the static composition over a mixing length. Assuming that the mixing length is of the order of the layer depth, the scaling becomes
\begin{equation}
     \Ro_0 = \Rey_0 \Ek \sim \frac{1}{r_i (\Rrhoi-1)} \frac{N^2}{\Omega^2}.
     \label{eq:Ro0_scaling}
\end{equation}
Figure~\ref{fig:Re0}(b) shows that the data \revision{(including the data from \citet{Monville2019} and \citet{Guervilly2022})} follow reasonably well the linear dependence on $N^2/\Omega^2$ predicted by the theoretical scaling for $N^2/\Omega^2\lesssim 10$. We determine empirically that $\Ro_0=10^{-2} N^2/\Omega^2$, meaning that the maximum value for $Ro_0$ is expected to be $0.1$ (for $N^2/\Omega^2 = 10$) irrespective of the Ekman number.
\revision{This scaling also provides a good fit to the data of \citet{Monville2019}, suggesting that mean flows driven by fingering convection in a rotating sphere or spherical shell saturate at similar amplitudes despite differences in model configuration, including boundary conditions and the radial dependence of the stratification.}
For $N^2/\Omega^2 \ge 100$, the Rossby number becomes a less meaningful measure of zonal flow amplitude as \revision{$\Ek \to 1$} in Series 1. Simulations at higher Rayleigh numbers and constant Ekman numbers, analogous to Series 2 but currently beyond computational reach, would likely reveal a decreasing trend of $\Ro_0$ with increasing $N^2$.

\subsection{Compositional transport}
\label{sec:Nusselt}

In this last result section, we study the mixing of the composition throughout the layer. We do not consider the convective heat transport since it is small compared with the conductive heat flux in fingering convection at low Prandtl numbers \citep{Garaud2018,Guervilly2022,Tassin2024}. We first look at the radial distribution of the convective flux, then its latitudinal dependence in the different regimes, and, finally, the global efficiency measured by a Nusselt number.

\subsubsection{Radial and latitudinal dependence}
The averaged convective and conductive fluxes of composition over a spherical surface $\mathcal{S}(r)$ are defined respectively as 
\begin{equation}
    \overline{\overline{F}}_{cv} = \Sc \int_{\mathcal{S}(r)} u_r \Cpert {\rm d}S, \quad  \overline{\overline{F}}_{cd} = -\int_{\mathcal{S}(r)} \frac{\partial \Cpert}{\partial r} {\rm d}S + F_s(r)
\end{equation}
where the double overbar denotes a spherical average and the static flux is $F_s(r) =- C'_s$.

\begin{figure}
	\centering 
    \includegraphics[width=\textwidth]{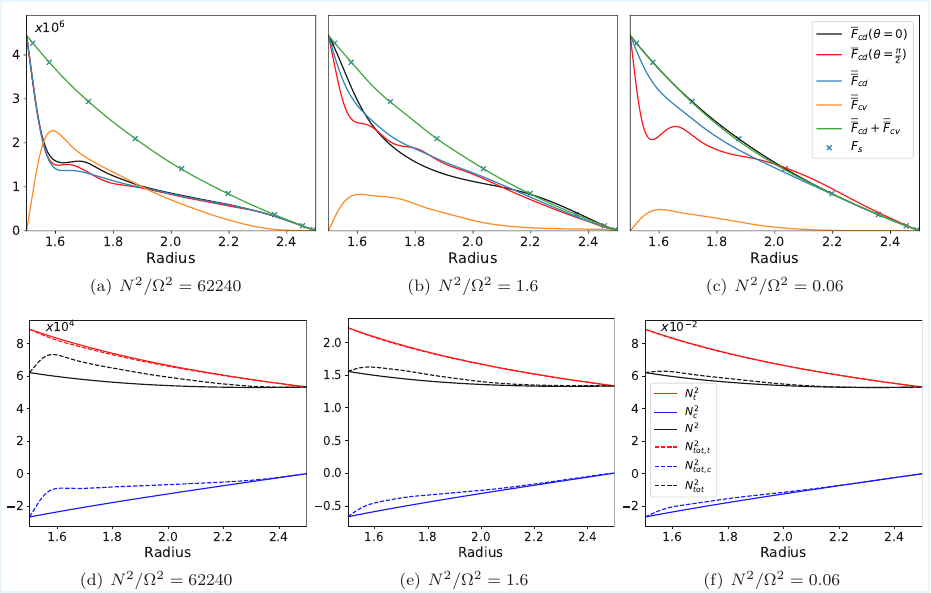}
	\caption{(a)-(c) Radial profiles of the convective ($F_{cv}$) and conductive ($F_{cd}$) fluxes of composition for (a) the weakly-rotating, (b) the intermediate, and 
    (c) the rapidly-rotating regimes from Series 1. The static flux in the absence of motion, $F_s$, is shown for comparison.
    Azimuthal averages are denoted by a single overbar and spherical averages by a double overbar. 
    \revision{(d)-(f) Radial profiles of the buoyancy frequency (normalised by $\Omega^2$) based on the static gradients (solid lines) for the density $(N^2)$, temperature $(N^2_{t})$ and composition $(N^2_{c})$ and based on the total gradients (dashed lines) for the density $(N^2_{tot})$, temperature $(N^2_{tot,t})$ and composition $(N^2_{tot,c})$.
    The profiles have been calculated from snapshots, but are representative of the statistically steady state.}} 	
    \label{fig:radial_flux}
\end{figure}

Figure~\ref{fig:radial_flux}\revision{(a)-(c)} shows the radial profiles of $\overline{\overline{F}}_{cv}$ and $\overline{\overline{F}}_{cd}$ for the three different dynamical regimes
for selected cases in Series 1. The sum of the convective and conductive fluxes equal the static flux at all radii, as expected from a steady state.
For $N^2/\Omega^2\gg 1$, the amplitude of the convective flux is comparable to the conductive flux in the lower domain of the layer where fingering convection is most vigorous.
The conductive fluxes in the polar regions ($\overline{F}_{cd}(\theta=0)$, where the single overbar denotes an azimuthal average) and in the equatorial region ($\overline{F}_{cd}(\theta=\pi/2)$) are similar, thereby confirming that there is no notable latitudinal difference when the rotational effects are weak.

For $N^2/\Omega^2\approx 1$, a reduction in convective flux is observed.
In the polar regions, where the bands are present, the conductive flux slightly diminishes at mid-radius compared with other latitudes, indicating that the bands enhance compositional mixing somewhat more effectively than the sheared fingers in the equatorial region.

For $N^2/\Omega^2\ll 1$, the convective flux is further reduced and significantly smaller than the conductive flux throughout the layer.
A large disparity between high and low latitudes is observed: only the region outside the tangent cylinder is convecting, so the conducting flux in the polar regions is superposed onto the static flux.  

\revision{To assess how the stratification is modified by fingering convection in the statistically steady state, Figure~\ref{fig:radial_flux}(d)-(f) shows radial profiles of the buoyancy frequency computed from the total gradients (\ie including both the perturbations and the static background) for density, temperature, and composition. These profiles are compared with the corresponding static profiles for the same three cases as Figure~\ref{fig:radial_flux}(a)-(c). The stabilising temperature stratification remains largely unchanged relative to the static state in all cases. However, the reduction of the destabilising compositional gradient leads to an increase in the overall stratification. This enhancement is most pronounced near the bottom of the layer, where it reaches up to 20\% in the weakly rotating case.}

\begin{figure}
	\centering 
	\includegraphics[height=5.5cm]{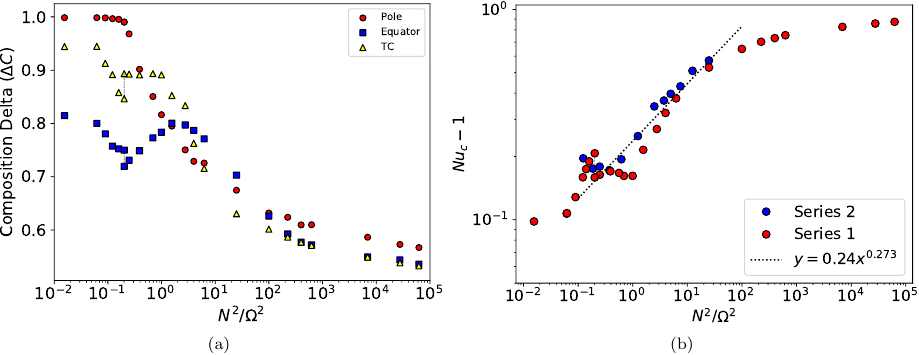}
	\caption{(a) Variation of $\Delta C$ at different latitudes as a function of $N^2/\Omega^2$ for Series 1.
    (b) Variation of $\Nu_c-1$ as a function of $N^2/\Omega^2$.
    The two solutions of the bistable case at $N^2/\Omega^2=0.2$ are linked by a dotted line.} 	
\label{fig:deltaC_Nu}
\end{figure}

To study systematically the latitudinal dependence of the transport efficiency as a function of $N^2/\Omega^2$, we measure the normalised composition difference across the layer at a given latitude
\begin{equation}
    \Delta C(\theta) = \frac{\overline{C}(r_o,\theta)-\overline{C}(r_i,\theta)}{C_s(r_o)-C_s(r_i)},
\end{equation}
where, as previously, the overbar denotes an azimuthal average. 
Because fixed compositional fluxes are imposed at both the inner and outer boundaries, we expect $\Delta C$ to decrease as the mixing efficiency increases.
Figure~\ref{fig:deltaC_Nu}(a) shows the values of $\Delta C$ averaged over a $\pm 5^{\circ}$ angle around three latitudes for Series 1: the north pole ($\theta = 0$), the equator ($\theta = \pi/2$), and the latitude where the tangent cylinder intersects the outer boundary ($\theta = 0.93$rad).

For $N^2/\Omega^2\ll1$, as expected, $\Delta C=1$ at the poles since there is initially no convection inside the tangent cylinder and transport occurs only in the equatorial region. The behaviour becomes notably non-monotonic in the rapidly-rotating regime at low latitudes, with a distinct discontinuity at $N^2/\Omega^2=0.2$, coinciding with the onset of convection in the polar regions. 
This discontinuity is associated with the reversal of the zonal flow and a pronounced increase in its amplitude (Figure~\ref{fig:Re0}(b)), indicating an intensification of radial shear.
Given the slight reduction in the poloidal Reynolds number following this transition (see Table~\ref{tab:Series1}), the observed increase of $\Delta C$ at low latitudes is likely a consequence of the enhanced shear, which may act to suppress convective transport.

For weak rotation ($N^2/\Omega^2>100$), $\Delta C$ attains similar values across latitudes, although the poles tend to exhibit slightly smaller values. This homogenisation of mixing is consistent with the gradual disappearance of zonal flows in this regime.

In the intermediate regime ($1\leq N^2/\Omega^2\leq 10$), $\Delta C$ at high latitudes is slightly smaller than at the equator, confirming moderately enhanced mixing associated with the bands inside the tangent cylinder.

\subsubsection{Nusselt number}
The efficiency of convective transport is commonly quantified by a global measure, the Nusselt number, which compares the total heat flux to the conductive heat flux at rest. In systems with fixed-flux boundary conditions, the Nusselt number can be determined from the spherically-averaged temperature difference across the layer \citep{Mound2017,Monville2019}. 
Thus, by analogy, we define the ``compositional" Nusselt number (sometimes called the Sherwood number) as
\begin{equation}
    \Nu_c = \frac{C_s(r_o) - C_s(r_i)}{\overline{\overline{C}}(r_o) - \overline{\overline{C}}(r_i)}.
\end{equation}
It is inversely related to the previously defined $\Delta C$, but provides a global measure across the entire layer, allowing its variation with control parameters to be compared with scaling predictions from the literature.

A power-law scaling of the form $\Nu_c-1= A \Ra_c^{1/3}$ (where the constant $A$ depends on $\Pran$, $\Le$, and $\Rrho$) has long been proposed based on dimensional arguments \citep{Turner1967}. This scaling has been observed in non-rotating, plane-layer experiments and numerical simulations for $\Pran>1$ \citep[\eg][]{Radko2000,Hage2010}. Extending the Rayleigh–B\'enard convection theory of \citet{Grossmann2000}, \citet{Yang2015} demonstrated that the $1/3$ law is expected in the asymptotic regime of large $\Ra_c$ for fingering convection. In contrast, for small $\Pran$ \revision{and in a triply periodic (unbounded) domain}, \citet{Brown2013} proposed an alternative scaling in which $Nu_c$ remains constant for fixed values of $\Pran$, $\Le$, and $\Rrho$ within the regime $\smallr \ll (1/\Le,\Pran) \ll1$ (noting that, although this regime is the closest for our simulations where $\Pran=0.3$, $\Le=10$ and $\smallr=0.26$, it is not strictly satisfied).  In the absence of rotation, \citet{Tassin2024} reported a best-fit scaling of $\Nu_c-1\sim\Ra_c^{0.27}$ from numerical simulations spanning a broad range of Prandtl numbers ($\Pran\in[0.03,7]$).  They attributed the deviation from the $1/3$ law to the non-negligible contribution of dissipation within the boundary layers, while the $1/3$ law in \citet{Grossmann2000} assumes bulk-dominated dissipation. In the rapidly-rotating regime with $N^2/\Omega^2<1$, \citet{Guervilly2022} found that the $1/3$ scaling remains a reasonable fit for constant density ratio, albeit with an additional dependence on the Ekman number.

We now examine how our simulations, which span a broad range of $N^2/\Omega^2$ at fixed density ratio, compare with the scaling behaviours discussed above. Figure~\ref{fig:deltaC_Nu}(b) shows the compositional Nusselt number for both Series as a function of $N^2/\Omega^2$.  Across all cases, the convective enhancement remains modest, $\Nu_c-1<1$, indicating that convective fluxes are relatively weak.
In the rapidly-rotating regime, the behaviour of $\Nu_c$ is non-monotonic, mirroring the behaviour of $\Delta C$ with a discontinuity at $N^2/\Omega^2=0.2$. The regional behaviours inside and outside the tangent cylinder act in opposition: the onset of fingering convection inside the tangent cylinder enhances convective transport, whereas the intensified shear in the equatorial region following the zonal flow reversal suppresses it. Overall, this results in a slight decrease in $\Nu_c$ at the transition.

For the range $0.2<N^2/\Omega^2<1$, $Nu_c$ remains approximately constant in Series 1, in line with the plateau observed in the poloidal Reynolds number over the same interval. 
In the intermediate regime, $1\leq N^2/\Omega^2 \leq 10$, $Nu_c$ begins to increase, mirroring the trend observed in Series 2.
As anticipated for $N^2/\Omega^2>100$, $Nu_c$ asymptotes to a plateau, indicating the transition to the \revision{$\Ek\to1$} dynamics in Series 1.
To facilitate comparison with previous studies, we fit a power-law scaling of the form $\Nu_c-1 \sim (N^2/\Omega^2)^{\alpha}$
over the range $1<N^2/\Omega^2<30$, despite noticeable scatter in the data.
The best fit across both Series yields an exponent $\alpha\approx 0.27$, closely matching the non-rotating scaling reported by \citet{Tassin2024}.
\revision{In the case of rotating fingering convection, the radial shear of the fingers due to the presence of the zonal flows indirectly modifies the transport efficiency and the intensification of shear with increasing $Ra_c$ (at least up to the peak in zonal flow amplitude at $N^2/\Omega^2 \approx 7.5$) may account for the observed shallower scaling relative to the classical 1/3 law.}

\section{Conclusions}
\label{sec:conclusion}

\subsection{\revision{Summary of the results}}
We have conducted numerical simulations of fingering convection in a thick spherical shell under the influence of rotation. Our study examined how the dynamics change as a function of the stratification strength $N$ measured relative to the rotation rate $\Omega$, by varying the Ekman number at fixed Rayleigh numbers (Series 1) or by varying the Rayleigh numbers at fixed Ekman number (Series 2). Throughout, the density ratio was fixed at $\Rrhoi = 3.33$ at the inner radius, as were the diffusivity ratios ($\Pran=0.3$ and $\Le = 10$). We found that the dynamics depend strongly on the parameter $N^2/\Omega^2$, with three distinct regimes emerging: a weakly-rotating regime for $N^2/\Omega^2 > 10$, a rapidly-rotating regime for $N^2/\Omega^2 < 1$, and an intermediate regime for $1 < N^2/\Omega^2 < 10$.

In all regimes, the primary fingering mode forms narrow, elongated fingers that align with the rotation axis when $N^2/\Omega^2 < 10$, and with gravity when $N^2/\Omega^2 > 10$. The fingers remain laminar, with local Reynolds numbers below unity. Their transverse scale $\lscale$ approximately follows the Stern scale $\ltheo = (|\Ra_t|/r_o)^{-1/4}$ and, for our fixed density ratio, is well described by $\lscale = 5\ltheo$, independent of the Ekman number. 
For Mercury's core with a stable layer of thickness $800$ km, the predicted transverse finger scales is approximately $50-0.5$ m for $N^2/\Omega^2 = 10^{-4}-10^4$
for the moderate density ratio considered here. 
In non-rotating fingering convection simulations, \citet{Tassin2024} found that the finger size depends only weakly on the density ratio. Consequently, primary fingers in planetary cores are expected to develop only on small spatial scales, on the order of 1 m. Note that an exception may occur close to the stability boundaries of fingering convection, especially when the stratification is extremely weak at low Rayleigh numbers. In this regime, columnar flows can develop with an azimuthal extent comparable to the layer thickness, as described by \citet{Monville2019,Mather2021,Guervilly2022}. In the limit $\Ek \ll 1$, the lower stability boundary occurs at $\Ra_c \approx 100 \Ek^{-1}$ \citep{Monville2019}, corresponding to $N^2 / \Omega^2 \approx 10^{-11}$ for Mercury.

Beyond the primary fingering mode, additional large-scale flows emerge depending on the regime.
In the range $1<N^2/\Omega^2<100$ and at sufficiently small Ekman numbers, we observe clustering of fingers in the upper domain, where the local density ratio is larger and fingering convection less vigorous. These clusters are associated with weak, large-scale density anomalies and surrounded by toroidal gyres driven by the Coriolis force. These toroidal gyres dominate the local dynamics. They differ from the toroidal jets reported by \citet{Tassin2024}, which flow along concentric shells about an arbitrary rotation axis.
The size of the density anomalies and associated gyres does not appear to vary significantly with the control parameters, although this cannot be established with certainty from our simulations given that they occur over a limited range of $\Ra_c$ and $\Ek$. Their typical azimuthal order is $m \approx 5-10$. If their size is indeed insensitive to these parameters, we may infer that such structures would have characteristic horizontal scales of approximately $500-1000$~km under Mercury’s core conditions.

In the intermediate regime ($1<N^2/\Omega^2<10$), fingering convection exhibits a novel feature: axisymmetric or spiraling bands of poloidal flow confined within the tangent cylinder. These bands form concentric cylindrical structures that are nearly invariant along the rotation axis and slowly drift equatorward. 
They are highly anisotropic, with latitudinal widths greater than the Stern scale (roughly twice the Stern scale, though their domain of existence is too narrow to confirm this reliably) but azimuthal extents far larger. 
They grow and saturate over viscous timescales, gradually superseding primary fingering, and contribute to compositional mixing slightly more efficiently than the surviving columnar fingers in the equatorial region. The bands are robustly found within $1\lesssim N^2/\Omega^2\lesssim10$ but do not persist outside this range. 

In the rapidly-rotating regime ($N^2/\Omega^2<1$), the mixing is laterally inhomogeneous, initially more vigorous near the equator. This generates latitudinal compositional gradients that are in a thermo-compositional wind balance, analogous to the thermal wind in single-component thermal convection \citep[\eg][]{Aubert2005}. 
The resulting radial shear is disruptive to fingering convection outside the tangent cylinder, weakening the mixing of composition in that region. As $N^2/\Omega^2$ increases, enhanced mixing at higher latitudes leads to a reversal of the latitudinal compositional gradient and, in turn, of the direction of the zonal flow, from retrograde to prograde. The zonal flow amplitude, quantified by the zonal Rossby number $\Ro_0$, scales linearly with $N^2/\Omega^2$ and reaches a maximum value of $\Ro_0 \approx 0.1$ at $N^2/\Omega^2=10$. For stronger stratification, the zonal flow gradually diminishes as rotation becomes less influential and the lateral inhomogeneity of the mixing fades. 

For rapid rotation at $N^2/\Omega^2 = 0.06$, we obtained a case in which an equatorially-antisymmetric and axisymmetric mode is favoured. This mode corresponds to hemispherical convection, where the fluid in the northern hemisphere is colder and richer in light elements than in the southern hemisphere. The associated equatorially-antisymmetric zonal flow (slower in the northern hemisphere) becomes dominant over the usual equatorially-symmetric zonal flow. This behaviour occurs only for this specific parameter set: the mode decays at smaller stratification and becomes subdominant at larger stratification. It closely resembles the linear mode identified by \citet{Monville2019} in a full-sphere geometry, indicating that it can also arise in a thick spherical shell.

\subsection{\revision{Interaction with magnetic fields}}
In most of the explored parameter space, we therefore find that fingering convection generates large-scale flows in addition to the primary small-scale fingers. The resulting large-scale dynamics is remarkably diverse, exhibiting axisymmetric bands of poloidal flows, finger clusters, toroidal gyres, hemispherical convection, and zonal flows. Although these flows generally have small local Reynolds numbers (except for the zonal flows), they may possess interesting dynamo properties or interact with ambient magnetic fields in ways that remain to be elucidated.

\revision{Few studies have investigated the interaction between fingering convection and ambient magnetic fields. In a plane-layer geometry and in the absence of rotation, \citet{Harrington2019} showed that the range of density ratios that are linearly unstable to fingering convection (equation~\ref{eq:range}) is unchanged by the presence of a vertical magnetic field. They found that the finger flows become increasingly rigid along the direction of the ambient field and that magnetised fingering convection saturates at higher amplitudes than its hydrodynamic counterpart, resulting in a substantial enhancement of convective transport. For simulations with Prandtl and Lewis numbers comparable to those considered here ($\Pran = 0.1$ and $\Le = 10$), the magnetically dominated regime is reached when the Alfvén velocity, $v_A = B_0/(\rho \mu_0)$ (where $B_0$ is the strength of the ambient magnetic field and $\mu_0$ the magnetic permeability) becomes comparable to the characteristic finger velocity, $v_f=\kappa_T/\ltheo$. This corresponds to magnetic field strengths satisfying $B_0>(\kappa_T \rho \mu_0)/\ltheo$. Applied to metallic planetary cores, this estimate suggests that magnetic fields exceeding $\sim 10$~nT may be sufficient to significantly enhance fingering convection, a condition met even by Mercury’s relatively weak magnetic field \citep{Anderson2012}. It should be noted, however, that the simulations of \citet{Harrington2019} were conducted at magnetic Prandtl numbers of order unity. As a result, this threshold may be overestimated for fluids with low magnetic Prandtl numbers \citep{Fraser2024}.}

One property of the small-scale fingers that may be relevant for magnetic interactions is their pronounced radial asymmetry in the regime weakly influenced by rotation ($N^2/\Omega^2 > 10$), characterized by intense, concentrated upflows surrounded by more diffuse downflows. This asymmetry could enable topological magnetic pumping, as first proposed by \citet{Drobyshevski1974} in the context of the solar interior. In the solar convection zone, compressible convection manifests as narrow downwelling lanes separating broader, weaker upflows; in this configuration, horizontal magnetic fields are expected to be carried downward by the convection \citep[\eg][]{Tobias2001}. In contrast, the radial asymmetry observed here is reversed, suggesting by analogy that magnetic flux might instead be pumped outward from the stable layer. However, it is unclear whether the degree of radial asymmetry associated with Boussinesq fingering convection would be sufficient for such magnetic flux pumping to operate effectively.

Of perhaps more direct relevance to the role of stable layers in shaping planetary magnetic fields is the spontaneous generation of strong zonal flows, even in the absence of imposed boundary heterogeneity. These zonal flows, and the associated shear, are often invoked as a mechanism to attenuate non-axisymmetric magnetic fields passing through a stable layer, thereby promoting axisymmetrisation of the surface magnetic field \citep{Stevenson1982}. According to \citet{Stevenson1982}’s linear analysis, achieving a dipole tilt below one degree at Mercury’s surface requires $\Ro_0 \gtrsim 10^{-4}$ for an $800$ km-thick stable layer (see discussion in \citet{Guervilly2022}). Our results suggest zonal flows sufficient to induce effective axisymmetrisation over $N^2/\Omega^2 \in [10^{-2}, 25]$. For stronger stratification, the flows may still generate adequate shear, but further simulations (particularly in Series 2 at fixed $\Ek = 10^{-4}$) are needed to confirm this. (Note that in Series 1, where \revision{$\Ek \to 1$}, the Rossby number becomes less meaningful for $N^2/\Omega^2 > 100$.)
\revision{To estimate a possible upper threshold at stronger stratification, we consider the last three data points of Series~2 in Figure~\ref{fig:Re0}(b), which exhibit a decreasing trend of $\Ro_0$ with increasing $N^2/\Omega^2$. A least-squares fit over the range $N^2/\Omega^2 \geq 7.5$ yields $\Ro_0=0.1 (N^2/\Omega^2)^{-0.45}$. Extrapolating this scaling suggests that $\Ro_0$ would decrease to $\sim 10^{-4}$ at $N^2/\Omega^2 \approx 5 \times 10^{6}$. While highly speculative, this estimate provides a tentative indication that zonal flows driven by fingering convection could provide adequate shear over a wide range of stratification strength.}

Ultimately, nonlinear magnetohydrodynamic simulations that include the feedback of the Lorentz force are required to properly test \revision{the interaction of rotating fingering convection with magnetic fields.}

\subsection{\revision{Limitations of the model}}
Finally, our results demonstrate that simulations extending over long integration times are essential to capture the rich dynamics of fingering convection: the equatorially-antisymmetric mode observed at $N^2 / \Omega^2 = 0.06$ grows and saturates only very slowly, becoming dominant after roughly 60 viscous timescales; similarly, the reversal of zonal flow direction at $N^2 / \Omega^2 = 0.2$ occurs only after several tens of viscous timescales; the bands of poloidal flow in the intermediate regime also require several viscous timescales to saturate. As a result, global simulations of fingering convection are computationally demanding, owing to the wide range of spatial and temporal scales involved. This limits our ability to approach realistic core conditions, particularly at low Ekman numbers. For example, in our exploration of the regime $N^2 / \Omega^2 > 1$, we were restricted to Ekman numbers $\Ek \geq 10^{-4}$, whereas in Mercury’s core $\Ek \approx 10^{-12}$. It is therefore difficult to determine whether the slow dynamics observed over viscous timescales in our simulations would remain relevant under planetary core conditions, where the characteristic viscous timescales at large scales exceed several billion years. 
These processes may plausibly evolve on shorter timescales in realistic core regimes, but confirming this would require computationally intensive simulations at lower Ekman numbers.

\revision{
In this study the Lewis number was fixed at $\Le=10$ due to computational constraints, whereas values in metallic planetary cores are expected to be closer to $\Le \sim 10^3$. On small scales, the use of a moderately large Lewis number is not expected to alter fundamentally the physics of fingering convection provided that $\Le \gg 1$ \citep{Stern2001,Radko2008}. However, the larger disparity between thermal and compositional diffusivities leads to enhanced vertical transport of composition, as reflected in the scaling laws for $\Nu_c$ derived by \citet{Traxler2011b,Brown2013} at $\Pran<1$ and \cite{Yang2020} at $\Pran>1$. On larger scales, the Lewis number also influences the emergence of thermohaline staircases: increasing $\Le$ promotes the $\gamma$-instability by shifting the minimum of the flux ratio $\gamma$ toward larger density ratios \citep{Schmitt1979,Radko2003}. Although this instability is expected to be inactive at low Prandtl number \citep{Traxler2011b}, staircase formation may still arise through alternative mechanisms \citep{Garaud2015}, which could themselves be sensitive to $\Le$.
As a first step toward exploring more realistic parameter regimes, we examined how increasing $\Le$ affects the poloidal flow bands identified in the intermediate regime, which constitute a novel feature of this work. We performed an additional simulation with $\Le=100$ (with $\Pran=0.1$ and $\Sc=10$) at $\Ek=10^{-4}$ and $N^2/\Omega^2=0.62$, adjusting the Rayleigh numbers to maintain the density ratio at $\Rrhoi=3.33$. We find that the poloidal bands persist under these conditions, indicating that this feature is robust to at least moderately larger $\Le$. Simulations at high Lewis numbers will be required to determine whether these results remain valid under planetary-relevant parameter regimes.} 

\revisiontwo{Lastly, our study neglects any dynamical interaction with the deeper convective core. We plan to extend this
work by modelling the coupling between convective and stable layers in the future.}

\section*{Acknowledgements}
This work was supported by the UK Science and Technology Facilities Council [ST/W001039/1; UKRI1201].
This research used the Rocket High Performance Computing service at Newcastle University and the DiRAC Data Intensive service (CSD3) at the University of Cambridge, managed by the University of Cambridge University Information Services on behalf of the STFC DiRAC HPC Facility (\url{www.dirac.ac.uk}). The DiRAC component of CSD3 at Cambridge was funded by BEIS, UKRI and STFC capital funding and STFC operations grants. DiRAC is part of the UKRI Digital Research Infrastructure.
For the purpose of open access, the authors have applied a Creative Commons Attribution (CC BY) licence to any Author Accepted Manuscript version arising from this submission.
The authors thank the two referees for suggestions that have improved the manuscript.

\section*{Data statement}
Data sets for this research are available on the Figshare powered
Newcastle University research data repository
(\url{https://data.ncl.ac.uk}) \citep{Gray2026data}. 
The numerical code (XSHELLS) used for this research is openly available at \url{https://nschaeff.bitbucket.io/xshells/}.

\appendix

\section{List of simulations}
\label{sec:appA}

\begin{table}[h!]
\centering
\begin{tabular}{ c c c c c c c c}
 \hline \hline
 $\Ek$ & $N^2/\Omega^2$ & \revision{$Ro_c$} & $(N_r, L_{max}, M_{max})$ & $\Rey$ & $\Rey_0$ & $Nu_c$& \revision{$\Delta t_{avg}$}\\
 \hline
 \hline
  $5\times 10^{-5}$ &0.016& \revision{0.13} & $(192, 120, 110)$ &2.0& 1.8  &1.098&\revision{5}\\
  \hline
  $1\times 10^{-4}$$^{\dagger}$  &0.062& \revision{0.26} & $(192, 120, 110)$ & 2.5 &3.7&1.108&\revision{5}\\
  \hline
  $1.2\times 10^{-4}$  & 0.090& \revision{0.31} & $(192, 120, 110)$ & 2.7 & 3.8 &1.128&\revision{10}\\
  \hline
  $1.4\times 10^{-4}$  & 0.12& \revision{0.36} & $(192, 120, 110)$ & 2.8 & 3.6&1.159&\revision{15}\\
  \hline
  $1.5\times 10^{-4}$  &0.14& \revision{0.39} & $(192, 120, 110)$ & 3.0 &3.2&1.174&\revision{12}\\
  \hline
  $1.6\times 10^{-4}$  & 0.16& \revision{0.41} & $(192, 120, 110)$ & 3.3 &2.6&1.190&\revision{12}\\
  \hline 
  $1.8\times 10^{-4}$$^{\ast}$  & 0.20& \revision{0.47} & $(192, 120, 110)$ & 3.5 &2.1&1.207&\revision{10}\\
  \hline
  $1.8\times 10^{-4}$$^{\ast\ast}$  & 0.20& \revision{0.47} & $(192, 120, 110)$ & 3.3 &6.8&1.159&\revision{12}\\
  \hline
  $2\times 10^{-4}$  &0.25& \revision{0.52} & $(192, 120, 110)$ & 3.0 &8.9&1.164&\revision{15}\\
  \hline
  $2.5\times 10^{-4}$  &0.39& \revision{0.65} & $(192, 120, 110)$ & 3.4 &13.5&1.170&\revision{10}\\
  \hline
  $3\times 10^{-4}$  &0.56& \revision{0.78} & $(192, 120, 110)$ & 3.3 &18.3&1.167&\revision{10}\\
  \hline
   $3.33\times 10^{-4}$  &0.69& \revision{0.86} & $(192, 120, 110)$ &  2.9 &21.5&1.162&\revision{5}\\ 
  \hline
  $4\times 10^{-4}$ & 1.0& \revision{1.0} & $(192, 120, 110)$ &3.1 & 27.2&1.162&\revision{5}\\
  \hline
  $5\times 10^{-4}$ &1.6& \revision{1.3} & $(192, 120, 110)$ & 3.4 &32.5&1.215&\revision{15}\\
  \hline
  $6.66\times 10^{-4}$  &2.8& \revision{1.7} & $(192, 120, 110)$ & 3.6 &44.1&1.270&\revision{5}\\
  \hline
  $8\times 10^{-4}$& 4.0& \revision{2.1} & $(192, 120, 110)$ & 3.8 &43.7&1.322&\revision{5}\\
  \hline
  $1\times 10^{-3}$ & 6.2& \revision{2.6}  & $(192, 120, 110)$ & 4.0 &42.3&1.378&\revision{5}\\
  \hline 
  $2\times 10^{-3}$  & 24.9 & \revision{5.2} & $(192, 120, 110)$ & 4.9 & 24.7& 1.530&\revision{5}\\
  \hline
  $4\times 10^{-3}$ &   99.6  & \revision{10} & $(192, 120, 110)$ &5.5& 13.1& 1.649&\revision{5}\\
  \hline
  $6\times 10^{-3}$  &224& \revision{16} & $(192, 120, 110)$ & 5.6 &8.6& 1.702&\revision{5}\\
  \hline
  $8\times 10^{-3}$ & 398& \revision{21} & $(192, 120, 110)$ & 5.7 &6.3& 1.731&\revision{5}\\
  \hline
  $1\times 10^{-2}$ & 623  & \revision{26}& $(192, 120, 110)$ & 5.7 & 5.2 & 1.754&\revision{5} \\ 
  \hline
  $3.33\times 10^{-2}$ &   6902  & \revision{86} & $(192, 120, 110)$ & 5.6 & 2.0 & 1.827&\revision{5} \\ 
  \hline
  $6.66\times 10^{-2}$  &27619& \revision{172} & $(192, 120, 110)$ &5.5& 1.3 & 1.754&\revision{5}\\
  \hline
  $1\times 10^{-1}$  & 62240    & \revision{258}& $(192, 120, 110)$ & 5.4 & 1.0 & 1.873&\revision{7}\\
  \hline \hline
  
\end{tabular}
\caption{Simulations from Series 1 with constant $Ra_c = 2\times10^7$, $\Ra_t=-\Ra_c/3$ and $\Le=10$. The numerical resolution is given by the number of radial points, $N_r$, and the truncation degree and order of the spherical harmonics decomposition, $L_{max}$ and $M_{max}$. 
\revision{$\Ro_c$ is the convective Rossby number calculated from the compositional Rayleigh number, $\Ro_c=\sqrt{\Ra_c \Ek^2/\Sc}$.}
$\Rey$ is the Reynolds number for the poloidal flow, $\Rey_0$ the Reynolds number for the zonal flow.
\revision{$\Delta t_{avg}$ is the run time used to compute the averaged diagnostics (in units of a viscous timescale).} 
$(^{\dagger})$ For $\Ek=10^{-4}$, the output quantities are calculated before the transition in the equatorial symmetry (see Table~\ref{tab:Series2}).
$(^{\ast})-(^{\ast \ast})$ The output quantities for the simulation at $\Ek=1.8\times 10^{-4}$ averaged before (after) the reversal of the direction of the zonal velocity are denoted by $^{\ast}$ ($^{\ast \ast}$ respectively).}
\label{tab:Series1}
\end{table}

\begin{table}[h!]
\centering
\begin{tabular}{ c c c c c c c c }
 \hline \hline
 $\Ra_c$ & $N^2/\Omega^2$ & \revision{$Ro_c$} & $(N_r, L_{max}, M_{max})$ & $\Rey$ & $\Rey_0$ & $Nu_c$& \revision{$\Delta t_{avg}$}\\
 \hline
 \hline
 $2\times10^7$$^{\dagger}$ &0.062& \revision{0.26} & $(192, 163, 163)$ &2.5&3.8&1.105& \revision{40}\\
 \hline
  $2\times10^7$$^{\dagger\dagger}$ &0.062& \revision{0.26} & $(192, 163, 163)$ &2.3&4.6&1.107& \revision{55}\\
 \hline
 $4\times10^7$ &0.12& \revision{0.36} & $(220, 163, 163)$ &4.0&3.5&1.196& \revision{20}\\
 \hline
 $6\times10^7$ &0.19& \revision{0.45} & $(250, 180, 160)$ &4.3&15.3&1.175& \revision{10}\\
 \hline
 $8\times10^7$ &0.25& \revision{0.52} & $(250, 180, 160)$ &4.7&24.2&1.179& \revision{10}\\
 \hline  
 $1.2\times10^8$ &0.37& \revision{0.63} & $(250, 180, 160)$ &5.1&41.6&1.172& \revision{10}\\
 \hline
 $2\times10^8$ &0.62& \revision{0.82} & $(300, 200, 180)$ &6.1&83.4&1.194& \revision{30} \\
 \hline 
 $4\times10^8$ &1.2& \revision{1.2} & $(300, 220, 220)$ &6.9&182.9&1.250& \revision{7}\\
 \hline
 $8\times10^8$ &2.5& \revision{1.6} & $(300, 220, 220)$ &8.5&303.8&1.346& \revision{6}\\
 \hline
 $1.2\times10^9$ &3.7& \revision{2.0} & $(300, 240, 220)$ &10.2&360.8&1.369& \revision{3} \\
 \hline
 $1.6\times10^9$ &5.0& \revision{2.3} & $(300, 240, 220)$ &11.5&391.0&1.397& \revision{6} \\
 \hline
  $2.4\times10^9$ &7.5& \revision{2.8} & $(300, 240, 220)$ &13.7&402.6&1.431& \revision{6}\\
 \hline
  $4\times10^9$ &12.4& \revision{3.7} & $(504, 380, 255)$ &17.9&309.2&1.511& \revision{3}\\
 \hline
  $8\times10^9$ &24.9& \revision{5.2} & $(608, 485, 340)$ &23.4&238.4&1.572& \revision{2}\\
 \hline \hline
\end{tabular}
\caption{Simulations from Series 2 with constant $\Ek = 10^{-4}$, $\Ra_t=-\Ra_c/3$ and $\Le=10$.
$(^{\dagger})-(^{\dagger \dagger})$ The output quantities for the simulation at $\Ra_c=2\times10^7$ averaged before (after) the transition in the flow equatorial symmetry are denoted by $^{\dagger}$ ($^{\dagger \dagger}$ respectively).}
\label{tab:Series2}
\end{table}


\newpage






\end{document}